    \documentclass[twocolumn,twocolappendix]{aastex63}
    \usepackage{
        amsmath,
        amssymb,
        newtxtext,
        newtxmath,
        graphicx,
        ae, aecompl,
        booktabs,
    }
    \usepackage[T1]{fontenc}

    \newcommand*{\rd}[2]{\frac{\mathrm{d}#1}{\mathrm{d}#2}}
    \newcommand*{\rtd}[2]{\frac{\mathrm{d}^2#1}{\mathrm{d}#2^2}}
    \newcommand*{\pd}[2]{\frac{\partial#1}{\partial#2}}
    \newcommand*{\rdil}[2]{\mathrm{d}#1 / \mathrm{d}#2}
    \newcommand*{\pdil}[2]{\partial#1 / \partial#2}

    \newcommand*{\abs}[1]{\left|#1\right|}
    
    \newcommand*{\bsmb}[1]{\boldsymbol{\mathbf{#1}}}
    \newcommand*{\uv}[1]{\hat{\bsmb{#1}}}
    \newcommand*{\p}[1]{\left(#1\right)}
    \newcommand*{\s}[1]{\left[#1\right]}
    \newcommand*{\z}[1]{\left\{#1\right\}}
    \DeclareMathOperator*{\argmin}{argmin}

    \colorlet{Corr}{red}
    \received{Apr 29, 2020}
    \revised{Sept 4, 2020}
    \accepted{Sept 8, 2020}
    \submitjournal{ApJ}

\shorttitle{Exoplanet Obliquities}
\shortauthors{Y.\ Su and D.\ Lai}

\begin{document}

\title{Dynamics of Colombo's Top: Generating Exoplanet Obliquities from
Planet-Disk Interactions}

\correspondingauthor{Yubo Su}
\email{yubosu@astro.cornell.edu}

\author[0000-0001-8283-3425]{Yubo Su}
\affiliation{Cornell Center for Astrophysics and Planetary Science, Department
of Astronomy, Cornell University, Ithaca, NY 14853, USA}

\author[0000-0002-1934-6250]{Dong Lai}
\affiliation{Cornell Center for Astrophysics and Planetary Science, Department
of Astronomy, Cornell University, Ithaca, NY 14853, USA}

\begin{abstract}

Large planetary spin-orbit misalignments (obliquities) may strongly influence
atmospheric circulation and tidal heating in the planet. A promising avenue to
generate obliquities is via spin-orbit resonances, where the spin and orbital
precession frequencies of the planet cross each other as the system evolves in
time. One such mechanism involves a dissipating (mass-losing) protoplanetary
disk that drives orbital precession of an interior planet. We study this
scenario analytically in this paper, and obtain the mapping between the general
initial spin orientation and the final obliquity. We show that (i) under
adiabatic evolution (i.e.\ the disk dissipates at a sufficiently slow rate), the
final planetary obliquity as a function of the initial spin orientation
bifurcates into distinct tracks governed by interactions with the resonance;
(ii) under nonadiabatic evolution, a broad range of obliquities can be excited.
We obtain analytical expressions for the final obliquities for various
regimes of parameter space. The dynamical system studied in this paper is an
example of ``Colombo's top'', and our analysis and results can be adapted to
other applications.

\end{abstract}


\section{Introduction}\label{s:intro}

\subsection{Colombo's Top}

A rotating planet is subjected to gravitational torque from its host star,
making its spin axis precess around its orbital (angular momentum) axis.  Now
suppose the orbital axis precesses around another fixed axis---such orbital
precession could arise from gravitational interactions with other masses in the
system (e.g.\ planets, external disks, or binary stellar companion). What is the
dynamics of the planetary spin axis?  How does the spin axis evolve as the spin
precession rate, the orbital precession rate, or their ratio, gradually changes
in time?

\citet{colombo1966} was the first to point out the importance of the above
simple model in the study of the obliquity (the angle between the spin and
orbital axes) of planets and satellites. Subsequent works
\citep{peale1969,peale1974possible,ward1975tidal,henrard1987} have revealed rich
dynamics of this model. With appropriate modification, this model can be used as
a basis for understanding the evolution of rotation axes of celestial bodies.
Indeed, many contemporary problems in planetary/exoplanetary dynamics can be
cast into a form analogous to this simple model or its variants
\citep[e.g.][]{ward2004I, fabrycky_otides, batygin2013magnetic, lai2014star,
anderson2018teeter,zanazzi2018planet}.

In this paper we present a systematic investigation on the secular evolution of
Colombo's top, starting from general initial conditions. Our study includes
several new analytical results that go beyond previous works. While our results
are general, we frame our study in the context of generating exoplanet
obliquities from planet-disk interaction with a dissipating disk.

\subsection{Planetary Obliquities from Planet-Disk Interaction}

It is well recognized that the obliquity of a planet may provide important clues
to its dynamical history. In the the Solar System, a wide range of planetary
obliquities are observed, from nearly zero for Mercury and $3.1^\circ$ for
Jupiter, to $23^\circ$ for Earth and $26.7^\circ$ for Saturn, to $98^\circ$ for
Uranus. Multiple giant impacts are traditionally invoked to generate the large
obliquities of ice giants \citep{original_gi, benz1989tilting,
korycansky1990one, morbidelli_gi}. For gas giants, obliquity excitation may be
achieved via spin-orbit resonances, where the spin and orbital precession
frequencies of the planet become commensurate as the system evolves
\citep{ward2004I, ward2004II, vokrouhlicky2015tilting}. Such resonances may also
play a role in generating the obliquities of Uranus and Neptune
\citep{hamilton_tilting_ice}. For terrestrial planets, multiple spin-orbit
resonances and their overlaps can make the obliquity vary chaotically over a
wide range \citep[e.g.][]{laskar1993chaotic, touma1993chaotic, correia2003long}

Obliquities of extrasolar planets are difficult to measure. So far only loose
constraints have been obtained for the obliquity of a faraway ($\gtrsim 50$~au)
planetary-mass companion \citep{bryan2020obliquity}. But there are prospects for
constraining exoplanetary obliquities in the coming years, such as using
high-resolution spectroscopy to obtain $v\sin i$ of the planet
\citep{snellen2014fast, bryan2018constraints} and using high-precision
photometry to measure asphericity of the planet \citep{seager2002constraining}.
Finite planetary obliquities have been indirectly inferred to explain the
peculiar thermal phase curves \citep[see e.g.][]{millholland_signatures,
ohno_infer_obl} and tidal dissipation in hot Jupiters
\citep{millholland_wasp12b} and in super-Earths
\citep{millholland2019obliquity}.

It is natural to imagine some of the mechanisms that generate planetary
obliquities in the Solar System may also operate in exoplanetary systems.
Recently, \citet{millholland_disk} studied the production of planet obliquities
via a spin-orbit resonance, where a dissipating protoplanetary disk causes
resonance capture and advection. In their work, a planet is accompanied by an
inclined exterior disk; as the disk gradually dissipates, the planetary
obliquity increases, reaching $90^\circ$ for what the authors characterize as
adiabatic resonance crossings.

The Millholland \& Batygin study assumes a negligible initial planetary
obliquity. This assumption is intuitive, since the planet attains its spin
angular momentum from the disk. But it may not always be satisfied. In
particular, the formation of rocky planets through planetesimal accretion can
lead to a wide range of obliquities, especially if the final spin is imparted by
a few large bodies \citep{dones1993does, lissauer1997accretion,
miguel2010planet}. Such ``stochastic'' accretion likely happened for terrestrial
planets in the Solar System. Giant impacts may have also played a role in the
formation of the close-in multiple-planet systems diskovered by the Kepler
satellite \citep[e.g.][]{inamdar2015formation, izidoro2017breaking}.

\subsection{Goal of This Paper}

In this paper, we consider a wide range of initial planetary obliquities in the
Millholland-Batygin dissipating disk scenario, and examine how the obliquity
evolves toward the ``final'' value as the exterior disk dissipates. We provide
an analytical framework for understanding the final planetary obliquity for
arbitrary initial spin-disk misalignment angles. We also consider various
dissipation timescales, and examine both ``adiabatic'' (slow disk dissipation)
and ``non-adiabatic'' evolution. We calibrate these analytical results with
numerical calculations. On the technical side, our paper
extends previous works \citep[such as][]{henrard1982, henrard1987,
millholland_disk} in several aspects. Two of our main results are: (i) a careful
accounting of the phase space area across sepratrix to analytically describe the
rich dynamics of adiabatic evolution, and (ii) using the concept of ``partial
adiabatic resonance advection'' to fully capture the dynamics in the
non-adiabatic limit.

It is important to note that while we focus on a specific scenario of
generating/modifying planetary obliquities from planet-disk interactions, our
analysis and results have a wide range of applicability. For example, a
dissipating disk is dynamically equivalent to an outward-migrating external
companion.

The paper is organized as follows. In Section~\ref{s:eq}, we review the relevant
spin-orbit dynamics and key concepts that are used in the remainder of the
paper. In Sections~\ref{s:ad} and~\ref{s:nonad}, we study the evolution of the
system when the disk dissipates on different timescales, from highly adiabatic
to nonadiabatic. Analytical results are presented to explain the numerical
results in both limits. We discuss the implications of our results in
Section~\ref{s:disk}. Our primary physical results consist of
Fig.~\ref{fig:ad_ensemble} in the adiabatic limit and
Fig.~\ref{fig:3_ensemble_05_15} in the nonadiabatic limit. Some detailed
calculations are relegated to the appendices, including a leading-order
estimate of the final planetary obliquities given small initial spin-disk
misalignment angles in Appendix~\ref{s:ad_approx}.

\section{Theory}\label{s:eq}

\subsection{Equations of Motion}

We consider a star of mass $M_\star$ hosting an oblate planet (mass $M_{\rm p}$,
radius $R_{\rm p}$ and spin angular frequency $\Omega_{\rm p}$) at semimajor
axis $a_{\rm p}$, and a protoplanetary disk of mass $M_{\rm d}$. For simplicity,
we treat the disk as a ring of radius $r_{\rm d}$, but it is simple to
generalize to a disk with finite extent \citep[see][]{millholland_disk}. Denote
$\bsmb{S}$ the spin angular momentum and $\bsmb{L}$ the orbital angular momentum of
the planet, and $\bsmb{L}_{\rm d}$ the angular momentum of the disk. The
corresponding unit vectors are $\uv{s} \equiv \bsmb{S} / S$, $\uv{l} \equiv \bsmb{L}
/ L$, and $\uv{l}_{\rm d} \equiv \bsmb{L}_{\rm d} / L_{\rm d}$.

The spin axis $\uv{s}$ of the planet tends to precess around its orbital
(angular momentum) axis $\uv{l}$, driven by the gravitational torque from the
host star acting on the planet's rotational bulge. On the other hand, $\uv{l}$
and the disk axis $\uv{l}_{\rm d}$ precess around each other due to
gravitational interactions. We assume $S \ll L \ll L_{\rm d}$, so $\uv{l}_{\rm
d}$ is nearly constant and $\uv{l}$ experiences negligible backreaction torque
from $\uv{s}$. The equations of motion for $\uv{s}$ and $\uv{l}$ in this limit
are \citep{anderson2018teeter}
\begin{align}
    \rd{\uv{s}}{t} &= \omega_{\rm sl} \p{\uv{s} \cdot \uv{l}}
            \p{\uv{s} \times \uv{l}}
        \equiv \alpha \p{\uv{s} \cdot \uv{l}}
            \p{\uv{s} \times \uv{l}},\label{eq:dsdt}\\
    \rd{\uv{l}}{t} &= \omega_{\rm ld}\p{\uv{l} \cdot \uv{l}_{\rm d}}
            \p{\uv{l} \times \uv{l}_{\rm d}}
        \equiv -g\p{\uv{l} \times \uv{l}_{\rm d}},
            \label{eq:dldt}
\end{align}
where
\begin{align}
    \omega_{\rm sl} &\equiv \frac{3GJ_2 M_pR_p^2 M_\star}{2a_p^3 I_p\Omega_p}
        = \frac{3k_{\rm qp}}{2k_{\rm p}} \frac{M_\star}{m_{\rm p}}
            \p{\frac{R_{\rm p}}{a_{\rm p}}}^3 \Omega_{\rm p},\label{eq:wsl}\\
    \omega_{\rm ld} &\equiv \frac{3M_{\rm d}}{4M_\star}\p{\frac{a_{\rm
            p}}{r_{\rm d}}}^3 n .\label{eq:wld}
\end{align}
In Eq.~\eqref{eq:wsl}, $I_p = k_p M_pR_p^2$ (with $k_p$ a constant) is the
moment of inertia and $J_2=k_{\rm qp}\Omega_p^2 (R_p^3/GM_p)$ (with $k_{qp}$ a
constant) the rotation-induced (dimensionless) quadrupole of the planet [for a
body with uniform density, $k_p=0.4, k_{qp}=0.5$; for giant planets, $k_p\simeq
0.25$ and $k_{qp}\simeq 0.17$ \citep[e.g.][]{lainey2016quantification}].
In other studies, $3k_{\rm qp} / 2 k_{\rm p}$ is often notated
as $k_2 / 2C$ \citep[e.g.][]{millholland_disk}. In Eq.~\eqref{eq:wld}, $n
\equiv \sqrt{GM_\star/a_{\rm p}^3}$ is the planet's orbital mean motion, and we
have assumed $r_d\gg a_p$ and included only the leading-order (quadrupole)
interaction between the planet and disk. We define three relative inclination
angles via
\begin{equation}
  \uv{s} \cdot \uv{l}\equiv \cos\theta,\quad
  \uv{s} \cdot \uv{l}_{\rm d}\equiv \cos\theta_{\rm sd},\quad
  \uv{l} \cdot \uv{l}_{\rm d}\equiv \cos I.
\end{equation}
In our model, $I$ is a constant.
Following standard notation \citep[e.g.][]{colombo1966,peale1969,ward2004I}, we
have defined $\alpha \equiv \omega_{\rm sl}$ and $g \equiv -\omega_{\rm ld}\cos I$.

We can combine Eqs.~\eqref{eq:dsdt} and~\eqref{eq:dldt} into a single equation
by transforming into a frame rotating about $\uv{l}_{\rm d}$ with frequency $g$.
In this frame, $\uv{l}_{\rm d}$ and $\uv{l}$ are fixed, and
$\uv{s}$ evolves as:
\begin{equation}
    \p{\rd{\uv{s}}{t}}_{\rm rot} = \alpha \p{\uv{s} \cdot \uv{l}}
            \p{\uv{s} \times \uv{l}}
        + g\p{\uv{s} \times \uv{l}_{\rm d}}.\label{eq:dsdt_rot}
\end{equation}

We define the dimensionless time $\tau$ as
\begin{equation}
    \tau \equiv \alpha t,
\end{equation}
and the frequency ratio $\eta$
\begin{align}
    \eta \equiv{}& -\frac{g}{\alpha}\nonumber\\
        ={}& 2.08 \p{\frac{k_{\rm p}}{k_{\rm qp}}}
            \p{\frac{\rho_{\rm p}}{\mathrm{g/cm}^3}}
            \p{\frac{M_{\rm d}}{0.01 M_{\odot}}}
            \p{\frac{a_{\rm p}}{5\;\mathrm{AU}}}^{9/2}\nonumber\\
        &\times
            \p{\frac{r_{\rm d}}{30 \;\mathrm{AU}}}^{-3}
            \p{\frac{M_\star}{M_{\odot}}}^{-3/2}
            \p{\frac{P_p}{10\;\mathrm{hrs}}}
            \cos I ,\label{eq:eta}
\end{align}
where $\rho_p = 3M_p/(4\pi R_p^3)$ and $P_p = 2\pi/\Omega_p$ is the planet's
rotation period. In Eq.~\eqref{eq:eta}, we have introduced the fiducial values
of variable parameters for the application considered in this paper.
Eq.~\eqref{eq:dsdt_rot} then becomes
\begin{equation}
    \p{\rd{\uv{s}}{\tau}}_{\rm rot} = \p{\uv{s} \cdot \uv{l}}
            \p{\uv{s} \times \uv{l}}
        - \eta\p{\uv{s} \times \uv{l}_{\rm d}}. \label{eq:dsdt_base}
\end{equation}

Throughout this paper, we consider $\alpha$ constant, but allow $g$ to vary in
time. In the dispersing disk scenario of \citet{millholland_disk}, $\abs{g}$
decreases in time due to the decreasing disk mass. We consider a simple
exponential decay model
\begin{equation}
    M_{\rm d}(t) = M_{\rm d}(0)e^{-t/t_{\rm d}},\label{eq:dmd_dt}
\end{equation}
with $t_{\rm d}$ constant. This implies
\begin{equation}
    \rd{\eta}{t} = -\eta /t_{\rm d},\;\; \mathrm{or}\;\;
    \rd{\eta}{\tau} = -\epsilon \eta\label{eq:deta_dt},
\end{equation}
where
\begin{align}
    \epsilon \equiv{}& \frac{1}{\alpha t_{\rm d}}\nonumber\\
    ={}& 0.106 \p{\frac{k_{\rm p}}{k_{\rm qp}}}
        \p{\frac{\rho_{\rm p}}{\mathrm{g/cm}^3}}
        \p{\frac{a_{\rm p}}{5\;\mathrm{AU}}}^3
        \p{\frac{P_{\rm p}}{10\;\mathrm{hrs}}}
        \p{\frac{t_{\rm d}}{\mathrm{Myr}}}^{-1}.\label{eq:eps_def}
\end{align}
Eqs.~\eqref{eq:dsdt_base} and~\eqref{eq:deta_dt} together constitute our system
of study.

In the next two subsections, we summarize the theoretical background relevant to
our analysis of the evolution of the system.

\subsection{Cassini States}\label{ss:cs}

\begin{figure}
    \centering
    \includegraphics[width=0.3\textwidth]{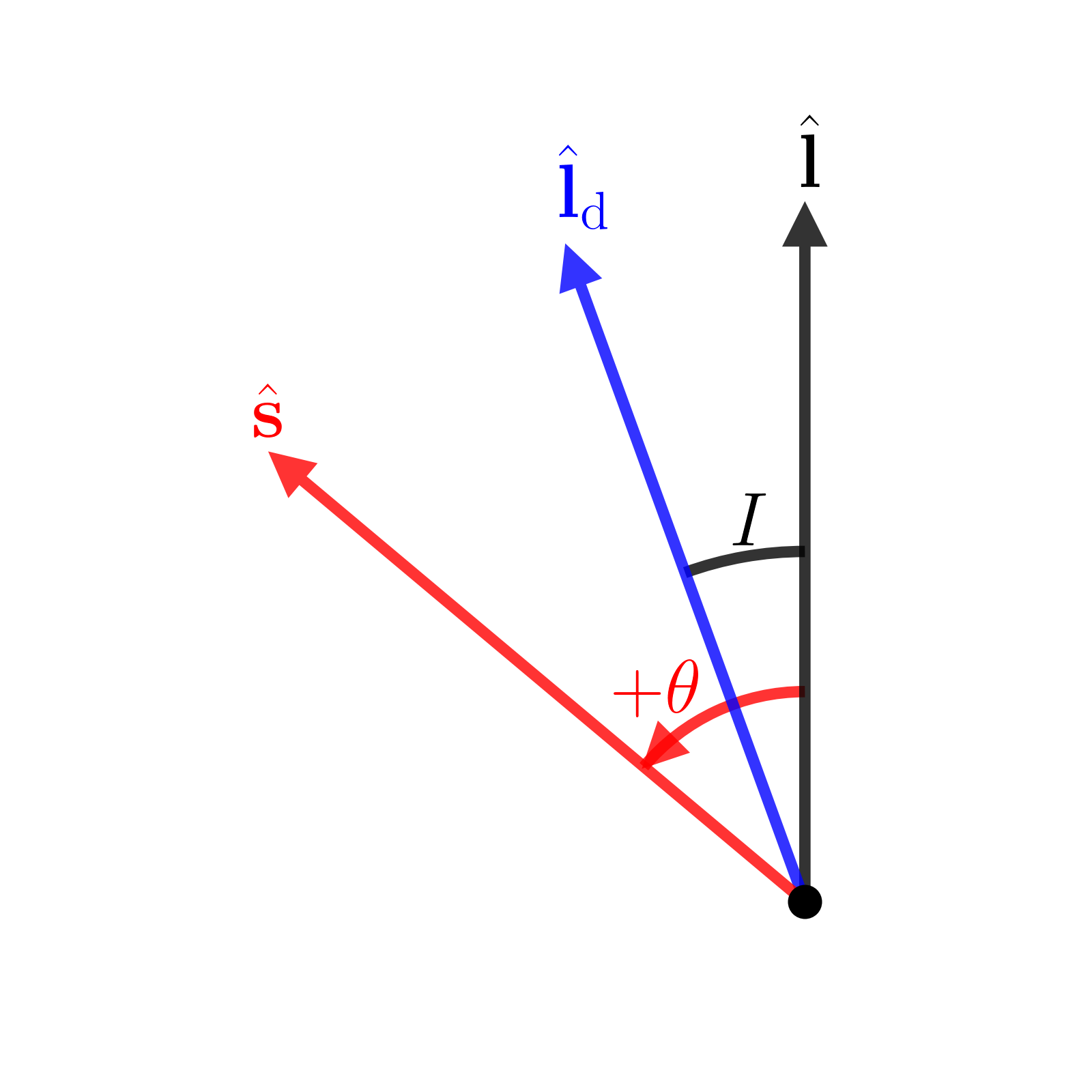}
    \caption{Definition of angles in the Cassini state configuration and the
    adopted sign convention for $\theta$. Traditionally, $\theta \in [-\pi,
    \pi]$.}\label{fig:cs_vecs}
\end{figure}

\begin{figure}
    \centering
    \includegraphics[width=0.47\textwidth]{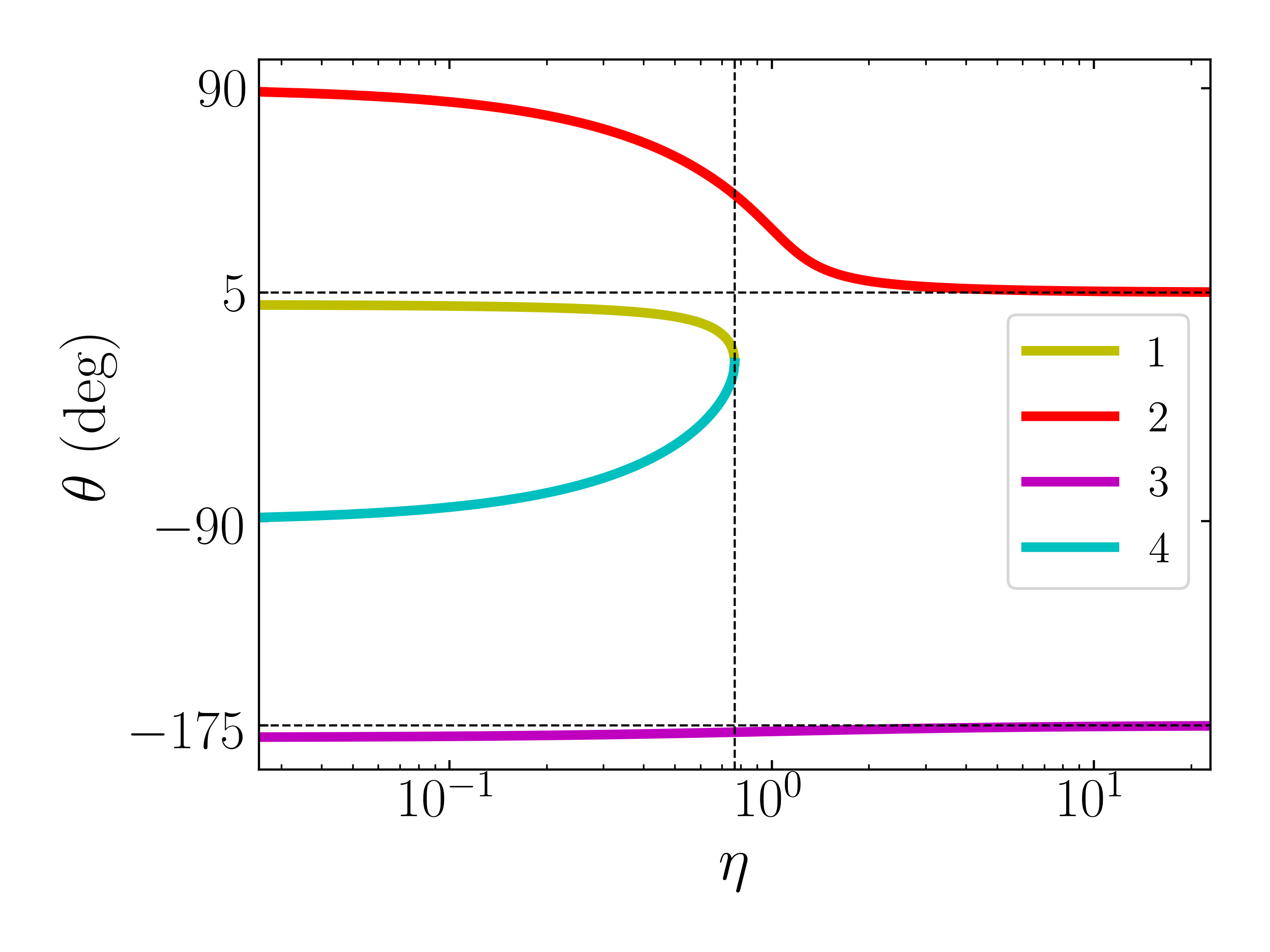}
    \caption{Cassini state obliquities as a function of $\eta$ for $I =
    5^\circ$. The thin vertical dashed line indicates $\eta_{\rm c}$ ($= 0.766$
    for $I = 5^\circ$), where CS1 and CS4 merge and annihilate, and the thin
    horizontal dashed lines indicate $\theta = I$ and $I - 180^\circ$, the
    asymptotic values for CSs 2 and 3 for $\eta \gg \eta_{\rm
    c}$.}\label{fig:cs_locs}
\end{figure}

Spin states satisfying $\p{\rdil{\uv{s}}{\tau}}_{\rm rot} = 0$ are referred to as
\emph{Cassini States} (CSs) \citep{colombo1966,peale1969}. They require that
$\uv{s}$, $\uv{l}$, and $\uv{l}_{\rm d}$ be coplanar. There are either two or
four CSs, depending on the value of $\eta$. They are specified by the obliquity
$\theta$ and the precessional phase of $\uv{s}$ around $\uv{l}$, denoted by
$\phi$. Following the standard convention and nomenclature (see
Figs.~\ref{fig:cs_vecs} and~\ref{fig:cs_locs}), CSs 1, 3, 4 have $\phi = 0$ and
$\theta < 0$, corresponding to $\uv{s}$ and $\uv{l}_{\rm d}$ being on opposite
sides of $\uv{l}$, while CS2 has $\phi = \pi$ and $\theta > 0$, corresponding to
$\uv{s}$ and $\uv{l}_{\rm d}$ being on the same side of $\uv{l}$. The CS
obliquity satisfies
\begin{equation}
    \sin \theta \cos \theta - \eta \sin\p{\theta - I} = 0.\label{eq:cs_zero}
\end{equation}
When $\eta < \eta_{\rm c}$, where
\begin{equation}
    \eta_{\rm c} \equiv \p{\sin^{2/3}\!I + \cos^{2/3}\!I}^{-3/2},\label{eq:etac}
\end{equation}
all four CSs exist, and when $\eta > \eta_{\rm c}$, only CSs 2, 3 exist. The CS
obliquities as a function of $\eta$ are shown in Fig.~\ref{fig:cs_locs}.

Of the four CSs, 1, 2, 3 are stable while 4 is unstable.
Appendix~\ref{s:local_dynamics} gives the libration frequencies and growth rates,
respectively, near these CSs.

\subsection{Separatrix}

\begin{figure*}
    \centering
    \includegraphics[width=0.8\textwidth]{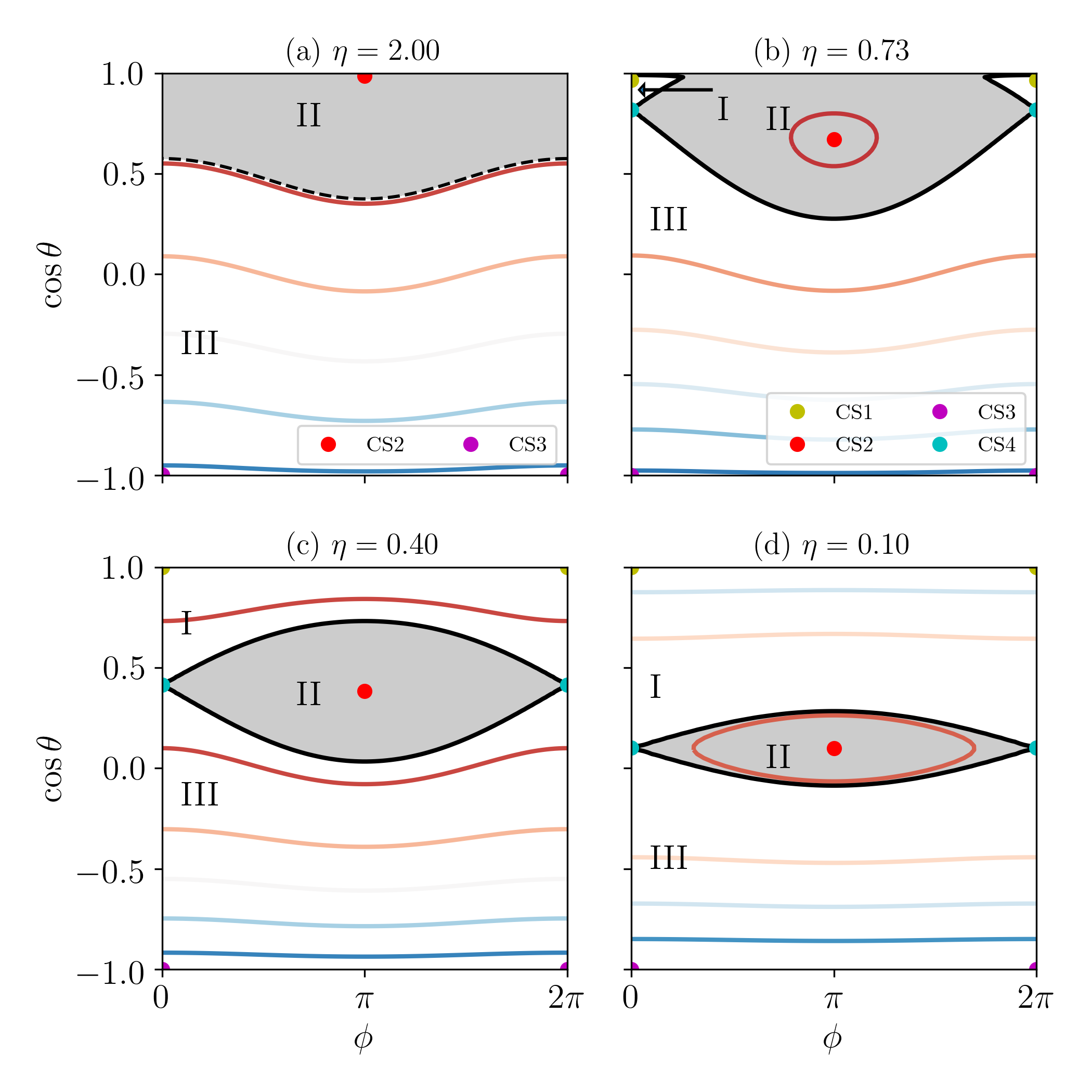}
    \caption{Level curves of $\mathcal{H}\p{\phi, \cos \theta}$
    [Eq.~\eqref{eq:H}] for $I = 5^\circ$, where warmer colors
    denote more positive values. The black solid line is the separatrix, which
    only exists for $\eta < \eta_{\rm c} = 0.766$. The three zones (I, II, III),
    divided by the separatrix, are labeled. The Cassini states are denoted by
    filled circles and have the same colors as in Fig.~\ref{fig:cs_locs}. The
    interior of the separatrix, shaded in grey, is formally only defined for
    $\eta < \eta_{\rm c}$, but we may identify the points in phase space that
    flow into zone II when evolved forward in time (decreasing $\eta$
    adiabatically); this is the shaded region in panel (a), bounded by the black
    dotted line.}\label{fig:eq_1contours}
\end{figure*}

The Hamiltonian (in the rotating frame) of the system is
\begin{align}
    \mathcal{H}\p{\phi, \cos \theta} &= -\frac{1}{2}\p{\uv{s} \cdot \uv{l}}^2
            + \eta \p{\uv{s} \cdot \uv{l}_{\rm d}}\nonumber\\
        &= -\frac{1}{2}\cos^2\theta
            + \eta \p{\cos \theta \cos I - \sin I \sin \theta \cos \phi}
                \label{eq:H}.
\end{align}
Here, $\phi$ and $\cos \theta$ are canonically conjugate variables. Trajectories
in the phase space $\p{\phi, \cos \theta}$ satisfy $H = $ constant (see
Fig.~\ref{fig:eq_1contours}).

When $\eta < \eta_{\rm c}$, CS4 exists and is a saddle point. The two
trajectories originating and ending at CS4 are the only two infinite-period
orbits in the phase space. Together, these two critical trajectories are
referred to as the \emph{separatrix} and divide phase space into three zones. In
Fig.~\ref{fig:eq_1contours}, we show the separatrix, the three zones, and their
relations to the CSs. Trajectories in zone II librate about CS2 while those in
zones I and III circulate.

\begin{figure}
    \centering
    \includegraphics[width=0.47\textwidth]{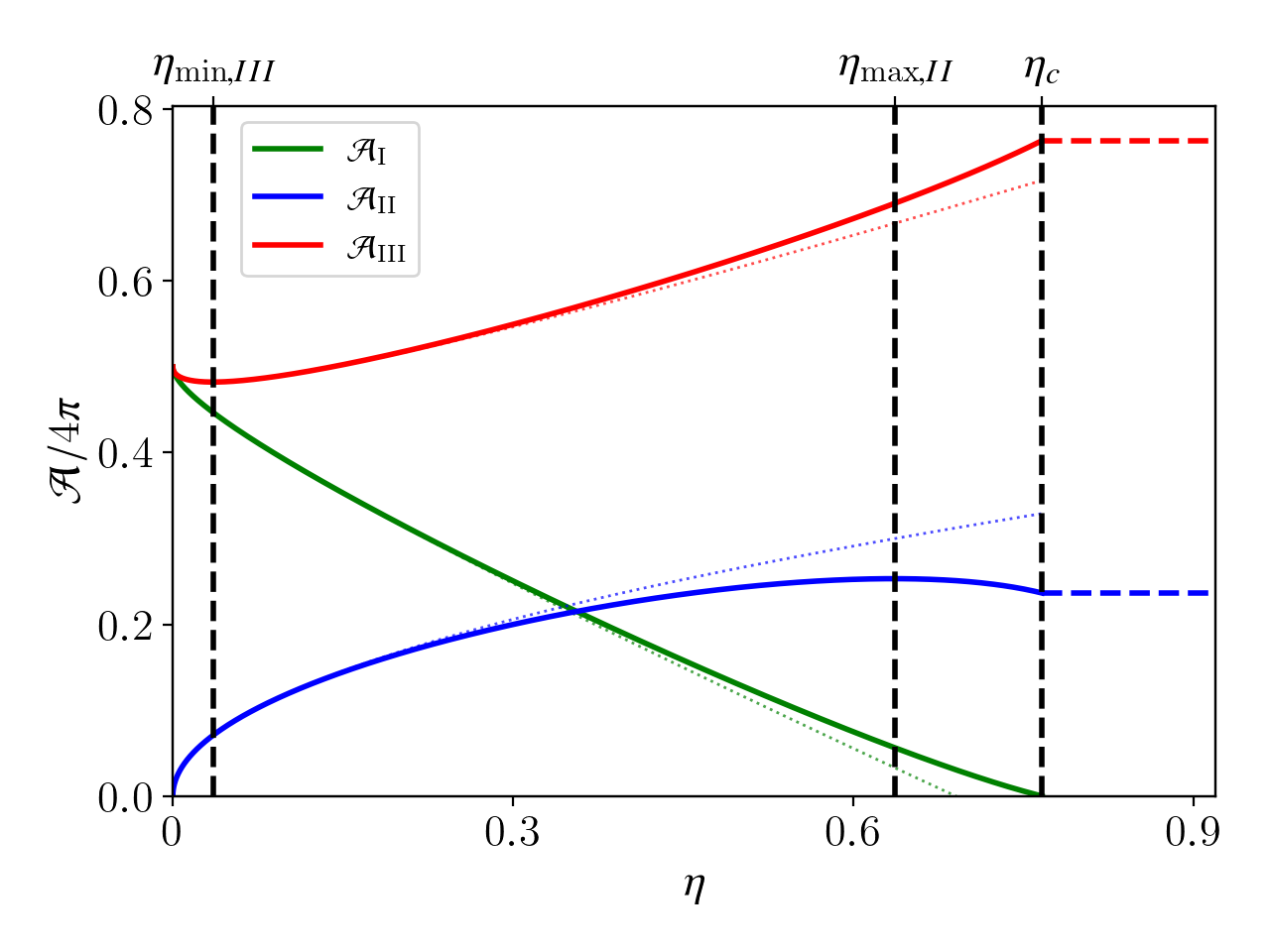}
    \caption{Plot of fractional areas of each of the zones
    $\mathcal{A}_{\rm j}(\eta) / 4\pi$ as given by Eqs.~\eqref{se:area_ward}
    for $I = 5^\circ$. The colored dotted lines correspond to
    small $\eta$ approximations used in Appendix~\ref{s:ad_approx}. The colored
    dashed lines for $\eta > \eta_{\rm c}$ are the effective values of
    $\mathcal{A}_{\rm II}, \mathcal{A}_{\rm III}$ for $\eta > \eta_{\rm c}$,
    denoting the points that would flow into either area under adiabatic
    decrease of $\eta$ from $\eta > \eta_{\rm c}$ (see the text). The vertical
    black dashed lines correspond to $\eta = \eta_{\rm c}$ [Eq.\eqref{eq:etac}]
    and the values of $\eta$ for which $\mathcal{A}_{\rm II}$ is maximized
    ($\eta_{\rm \max, II}$) and for which $\mathcal{A}_{\rm III}$ is minimized
    ($\eta_{\rm \min, III}$, Eq.~\eqref{eq:eta_minIII}).}\label{fig:eq_areas}
\end{figure}

Since $\p{\phi, \cos \theta}$ are canonically conjugate, the integral $\oint
\cos \theta\;\mathrm{d}\phi$ along a trajectory is an adiabatic invariant (see
Section~\ref{s:ad}). The unsigned areas $\p{\abs{\int \cos \theta
\;\mathrm{d}\phi}}$ of the three zones (as defined in
Fig.~\ref{fig:eq_1contours}) can be computed analytically. If we define
\begin{subequations}
    \begin{align}
        z_0 &= \eta\cos I, &
        \chi &= \sqrt{-\frac{\tan^3\theta_4}{\tan I} - 1},\\
        \rho &= \chi \frac{\sin^2 \theta_4\cos \theta_4}{
            \chi^2 \cos^2\theta_4 + 1},&
        T &= 2\chi \frac{\cos \theta_4}{
            \chi^2 \cos^2\theta_4 - 1},
    \end{align}
\end{subequations}
then the areas for $\eta < \eta_{\rm c}$ are given by
\citep{ward2004I}
\begin{subequations}\label{se:area_ward}
    \begin{align}
        \mathcal{A}_{\rm I} &= 2\pi\p{1 - z_0} - \frac{\mathcal{A}_{\rm II}}{2},
            \label{eq:A1}\\
        \mathcal{A}_{\rm II} &= 8\rho + 4\arctan T - 8z_0 \arctan
            \frac{1}{\chi}, \label{eq:A2}\\
        \mathcal{A}_{\rm III} &= 2\pi\p{1 + z_0} - \frac{\mathcal{A}_{\rm
            II}}{2}\label{eq:A3}.
    \end{align}
\end{subequations}
These are plotted as a function of $\eta$ in Fig.~\ref{fig:eq_areas}. While the
zones are not formally defined for $\eta > \eta_{\rm c}$ since the separatrix
disappears, a natural extension exists: evolve an initial phase space point $p$
under adiabatic decrease of $\eta$ until the separatrix appears at $\eta =
\eta_{\rm c}$, then identify $p$ with the zone it is in at $\eta_{\rm c}$. Since
phase space area is conserved under adiabatic evolution, this extension implies
$\mathcal{A}_{\rm j}\p{\eta > \eta_{\rm c}} = \mathcal{A}_{\rm j}(\eta_{\rm
c})$. The boundary between these extended zones is denoted by the dashed black
line in panel (a) of Fig.~\ref{fig:eq_1contours}, where no separatrix exists.

\section{Adiabatic Evolution}\label{s:ad}

In this section, we study the evolution of the planetary obliquity $\theta$ when
the parameter $\eta$ [or the disk mass $M_{\rm d}$; see Eqs.~\eqref{eq:eta}
and~\eqref{eq:deta_dt}] decreases sufficiently slowly that the evolution is
adiabatic. Intuitively, this requires the disk evolution time
$t_{\rm d}$ [Eq.~\eqref{eq:dmd_dt}] be much larger than the spin precession
period, $2\pi / \alpha$, i.e.\ $\epsilon = 1 / (\alpha t_{\rm d}) \ll 1/
(2\pi)$.

More rigorously, adiabaticity requires $t_{\rm d}$ be much larger than all
timescales of the dynamical system governed by the Hamiltonian
[Eq.~\eqref{eq:H}]. This is of course not possible in all cases, as the motion
along the separatrix has an infinite period. In practice, as $\eta$ evolves, the
system only crosses the separatrix once or twice, while it spends many orbits
inside one of the three zones and far from the separatrix. Thus, a \emph{weak
adiabaticity criterion} is that, for all equilibria/fixed points, the local
circulation/libration periods are much shorter than the timescale for the motion
of the equilibria due to changing $\eta$. If this criterion is satisfied, then
the system will evolve adiabatically for most of its evolution save one or two
separatrix crossings.

As shown in Appendix~\ref{ss:eigens}, libration about CS2 is slower than that
about CS1 or CS3. As such, it has the smallest characteristic frequency in the
system. The weak adiabaticity criterion is equivalent to
requiring that, at all times other than separatrix crossing, the obliquity of
CS2 ($\theta_2$) evolve over a longer timescale than the local libration period
about CS2, i.e.
\begin{equation}
    \abs{\rd{\theta_2}{\tau}} \ll
        \frac{\omega_{\rm lib}}{2\pi},\label{eq:ad_base}
\end{equation}
where
\begin{equation}
    \omega_{\rm lib} = \sqrt{\eta\sin I \sin \theta_2
        \p{1 + \eta \sin I \csc^3 \theta_2}},\label{eq:w_lib}
\end{equation}
is the libration frequency about CS2 for a given $\eta$
(Appendix~\ref{ss:eigens}). This formula differs from that given in
\citet{millholland_disk}, where the $\csc^3\!\theta_2$ term is neglected and the
square root is missing\footnote{The missing $\csc^3\!\theta_2$ term can be
traced to a $\theta \gg I$ approximation made in Eq.~(3) of \citet{ward2004II}.
Since $\theta_2 \sim I$ for $\eta \gg 1$ (Fig.~\ref{fig:cs_locs}), this
approximation is not always valid.}. Differentiating Eq.~\eqref{eq:cs_zero}
gives
\begin{equation}
    \rd{\theta_2}{\tau} = -\epsilon \frac{\eta \sin\p{\theta_2 - I}}{
        \cos\p{2\theta_2} - \eta \cos\p{\theta_2 - I}},\label{eq:dq2dt}
\end{equation}
where $\epsilon = -\rdil{(\ln \eta)}{\tau}$ [Eq.~\eqref{eq:eps_def}].
Eq.~\eqref{eq:ad_base} is most constraining at $\eta \sim 1$, i.e.\ it will be
satisfied for all $\eta$ if it is satisfied near $\eta \sim 1$, where
$\abs{\rdil{\theta_2}{\tau}} \sim \epsilon$. Thus, weak adiabiticity requires
\begin{equation}
    \epsilon \ll \epsilon_{\rm c}
        \equiv \p{\frac{\omega_{\rm lib}}{2\pi}}_{\eta = 1}
            \simeq \frac{1}{2\pi\sqrt{2}}\sqrt{\sin I\p{1 + 8\sin I}},
            \label{eq:ad_constr}
\end{equation}
where in the last equality we have used $\sin \theta_{\rm 2} \simeq 1/2$ at
$\eta = 1$ (e.g.\ when $I = 5^\circ$ and $\eta = 1$, $\theta_{\rm 2} \approx
31^\circ$). For $I = 5^\circ$, we obtain $\epsilon_{\rm c} \approx 0.0433$.
Since our criterion is only a weak condition for adiabaticity, we use $\epsilon
= 3 \times 10^{-4}$ in our ``adiabatic'' calculations below. We explore the
consequences of nonadiabatic evolution in Section~\ref{s:nonad}.

\subsection{Adiabatic Evolution Outcomes}\label{ss:ad_ensemble}

We consider the evolution of a system with arbitrary initial spin-disk
misalignment angle $\theta_{\rm sd, i}$ and initial $\eta_{\rm i} \gg 1$. We are
interested in the final spin obliquities $\theta_{\rm f}$ after $\eta$ gradually
decreases to $\eta_{\rm f} \ll 1$ (i.e.\ after the disk has dissipated to a
negligible mass). Note that when $\eta_{\rm i} \gg 1$, $\uv{l}$ precesses
around $\uv{l}_{\rm d}$ much faster than the spin-orbit precession
($\abs{\omega_{\rm ld}} \gg \abs{\omega_{\rm sl}}$), and the spin obliquity
$\theta$ varies rapidly. It is thus more appropriate to use $\theta_{\rm sd, i}$
rather than $\theta$ to specify the initial spin orientation. We explore the
entire range $\theta_{\rm sd, i} \in [0, \pi]$ and choose $\epsilon = 3 \times
10^{-4}$ (see above).

\begin{figure}
    \centering
    \includegraphics[width=0.47\textwidth]{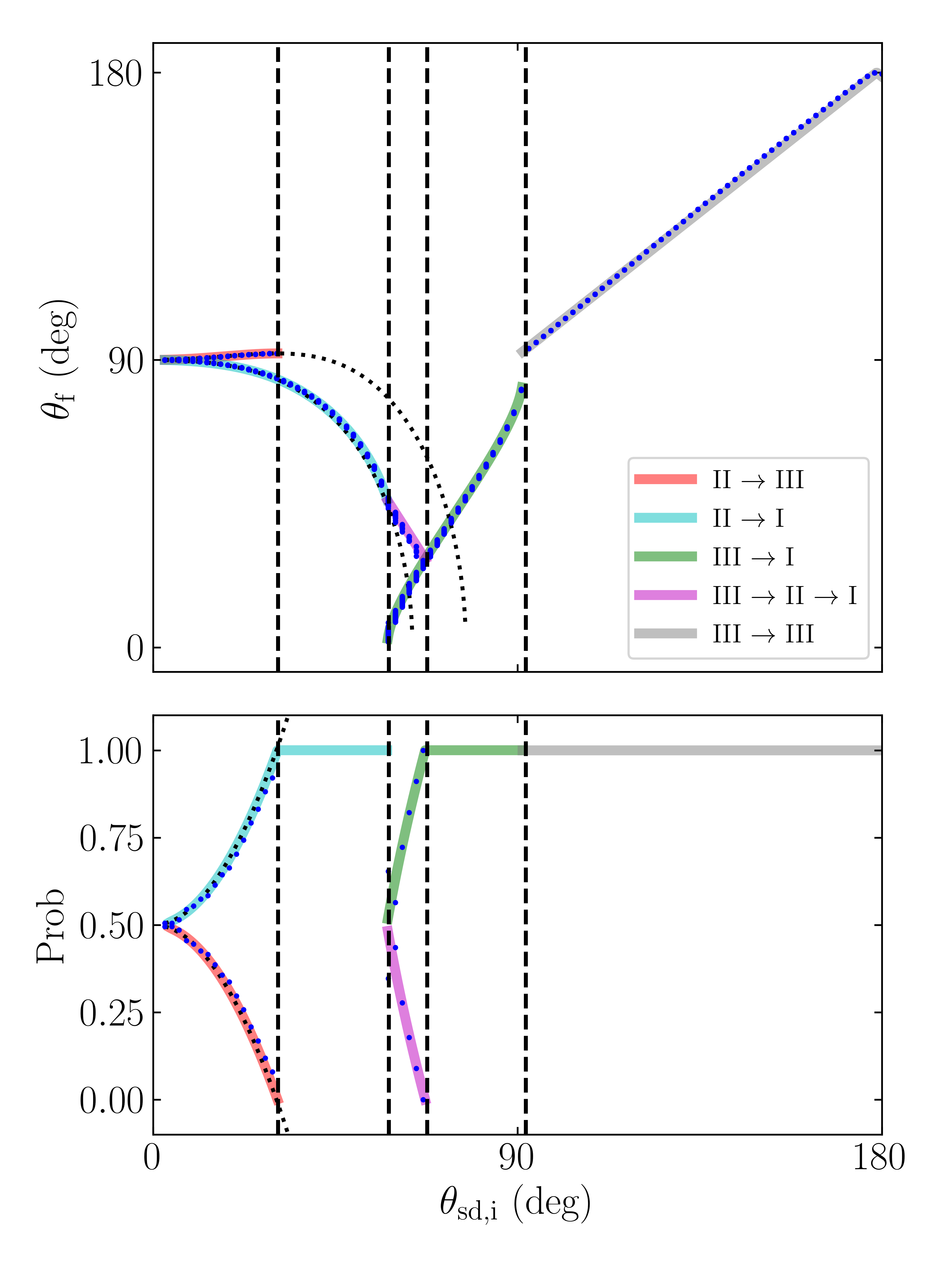}
    \caption{Top: The final spin obliquity $\theta_{\rm f}$ as a function of the
    initial spin-disk misalignment angle $\theta_{\rm sd, i}$ for systems
    evolving from initial $\eta_{\rm i} \gg 1$ to $\eta_{\rm f} \ll 1$,
    for $I = 5^\circ$. The blue dots are results of numerical
    calculations (Section~\ref{ss:ad_ensemble}), and the colored tracks are
    semi-analytical results (Section~\ref{ss:zone_transitions}). Bottom: The
    probabilities of different outcomes. Where a particular $\theta_{\rm sd, i}$
    corresponds to multiple tracks, the system evolves probabilistically. The
    track that a particular system evolves along in a numerical simulation can
    be measured by examining its final obliquity. The dots represent the
    inferred probabilities from measured final obliquities in our simulations,
    while the colored tracks denote the semi-analytic probability of the system
    evolving along each track. There are five regimes of $\theta_{\rm sd, i}$
    values for which different tracks are accessible. In both plots, the
    vertical dashed black lines denote semi-analytical calculations of the
    boundaries of these regimes (see Section~\ref{ss:zone_transitions}), while
    the black dotted lines represent analytical approximations valid in the
    small-$\theta_{\rm sd, i}$ limit (see
    Appendix~\ref{s:ad_approx}).}\label{fig:ad_ensemble}
\end{figure}

\begin{figure}
    \centering
    \includegraphics[width=0.47\textwidth]{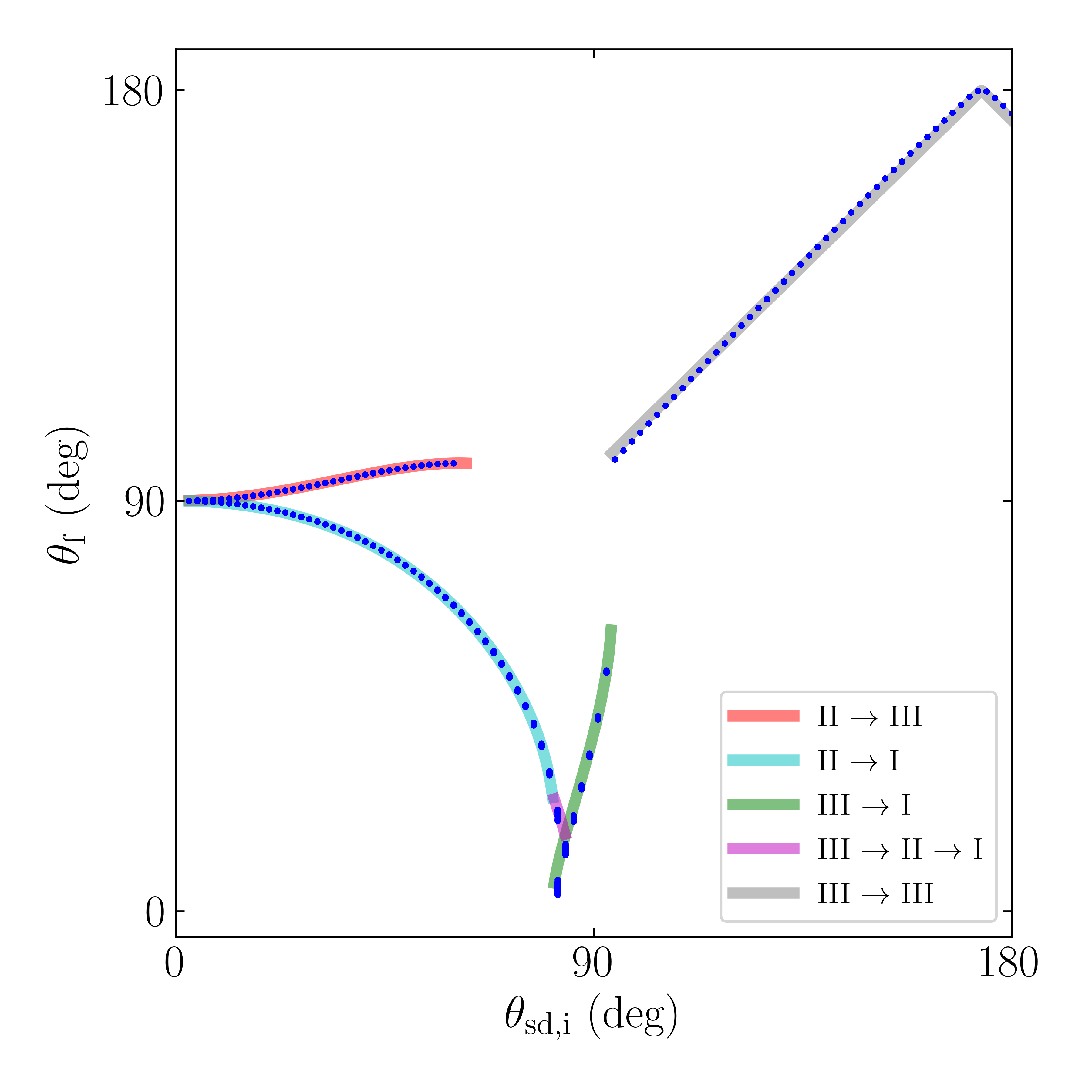}
    \caption{Same as the top panel of Fig.~\ref{fig:ad_ensemble} but for $I
    = 20^\circ$ and with fewer annotations.}\label{fig:3_ensemble_20_35}
\end{figure}

To obtain the distribution of the final obliquities $\theta_{\rm f}$, we evenly
sample $101$ values of $\theta_{\rm sd, i}$, and for each $\theta_{\rm sd, i}$
value, we pick $101$ evenly spaced orientations of $\uv{s}$ approximately from
the ring of initial conditions having angular distance $\theta_{\rm sd, i}$ to
$\uv{l}_{\rm d}$%
\footnote{The actual procedure we adopt to choose the initial conditions is the
natural extension of this description to finite $\eta_{\rm i}$. Note that the
center of libration of $\uv{s}$ is CS2, which, since $\eta_{\rm i}$ is finite,
is different from $\uv{l}_{\rm d}$. Furthermore, the libration is not exactly
circular. As a result, the libration trajectories for initial conditions on the
circular ring of points having angular distance $\theta_{\rm sd, i}$ from
$\uv{l}_{\rm d}$ are not the same and will each enclose slightly different
initial phase space areas $A_{\rm i}$. Since our analytical theory assumes exact
conservation of the initially enclosed phase space area $A_{\rm i}$ for each
$\theta_{\rm sd, i}$ (see Section~\ref{ss:zone_transitions}), this diskrepancy
introduces an extra deviation from the analytical prediction. To guarantee all
points for a particular $\theta_{\rm sd, i}$ have the same $A_{\rm i}$, we
instead choose initial conditions on the libration cycle going through
$\p{\theta_2 + \theta_{\rm sd, i}, \phi_2}$ [where $\p{\theta_2, \phi_2}$ are
the coordinates of CS2]. This ensures that all initial conditions for a given
$\theta_{\rm sd, i}$ enclose the same initial $A_{\rm i}$. As $\eta_{\rm i} \to
\infty$, this procedure generates initial conditions on the ring having angular
distance $\theta_{\rm sd, i}$ to $\uv{l}_{\rm d}$, recovering the procedure
given in the text.}.
To be concrete, we choose $\eta_{\rm i} = 10\eta_{\rm c}$ where $\eta_{\rm c}$
is given by Eq.~\eqref{eq:etac} and evolve Eqs.~\eqref{eq:dsdt_base}
and~\eqref{eq:deta_dt} until $\eta$ reaches its final value $10^{-5}$. At such a
small $\eta$, $\uv{s}$ is strongly coupled to $\uv{l}$ and the final obliquity
$\theta_{\rm f}$ is frozen. The mapping between $\theta_{\rm sd, i}$ and
$\theta_{\rm f}$ is our primary result, and is shown for $I = 5^\circ$,
$10^\circ$, and $20^\circ$ in Figs.~\ref{fig:ad_ensemble}
and~\ref{fig:3_ensemble_20_35} respectively. The blue dots represent the results
of the numerical calculation. The colored tracks are calculated
semi-analytically using the method discussed in the following subsection.


\subsection{Analytical Theory for Adiabatic Evolution}\label{ss:zone_transitions}

The evolutionary tracks that govern the $\theta_{\rm f}$-$\theta_{\rm sd, i}$
mapping correspond to various sequences of separatrix crossings. They can be
understood using the principle of adiabatic invariance, combined with (i) how
the enclosed phase space area by the trajectory evolves across each separatrix
crossing, and (ii) the associated probabilities with each separatrix crossing.

\subsubsection{Governing Principle: Evolution of Enclosed Phase Space
Area}\label{sss:a_evo}

First, we consider how the enclosed phase space area by a trajectory evolves
over time. In the absence of separatrix encounters, the enclosed phase space
area $\oint \cos\theta \;\mathrm{d}\phi$ is an adiabatic invariant. We adopt
convention where
\begin{equation}
    A \equiv \oint \p{1 - \cos \theta}\;\mathrm{d}\phi.\label{eq:a_oint}
\end{equation}
Note that $A$ can be negative when $\rdil{\phi}{t} < 0$, unlike
the unsigned areas $\mathcal{A}_{\rm i}$ [Eqs.~\eqref{se:area_ward}] which are
positive by definition. This definition
of $A$ has two advantages: (i) it is continuous across transitions from
circulating to librating that cross the North pole ($\cos \theta = 1$), and (ii)
the areas of the three zones are equal in absolute value to the expressions
given in Eqs.~\eqref{se:area_ward}. The path over which the integral is taken is
either a libration or circulation cycle. When $\eta_{\rm i} \gg 1$, trajectories
librate about $\uv{l}_{\rm d}$ with constant $\theta_{\rm sd}$, meaning they
enclose initial phase space area
\begin{equation}
    A_{\rm i} = 2\pi\p{1 - \cos \theta_{\rm sd, i}}.\label{eq:ai_qsd}
\end{equation}
Complications arise when considering finite $\eta_{\rm i}$, as trajectories near
CS2 or CS3 librate about these equilibria, rather than $\uv{l}_{\rm d}$, and
Eq.~\eqref{eq:ai_qsd} is no longer exact. In practice, Eq.~\eqref{eq:ai_qsd}
holds very well when defining $\theta_{\rm sd, i}$ as the angular distance to
CS2; an exception is discussed in Section~\ref{sss:evol_traj}.

Beginning at the last separatrix crossing, the final enclosed phase space area
$A_{\rm f}$ will be conserved for all time. As $\eta \to 0$, trajectories
circulate about $\uv{l}$ at constant obliquity $\theta_{\rm f}$, related to
$A_{\rm f}$ by
\begin{equation}
    2\pi\p{1 - \cos \theta_{\rm f}} = A_{\rm f}. \label{eq:qfaf}
\end{equation}

The enclosed phase space area is not conserved when the trajectory encounters
the separatrix. However, the change is easily understood \citep{henrard1982}. In
essence, when the trajectory crosses the separatrix, it continues to evolve
adjacent to the separatrix. So if a separatrix crossing results in a zone I
trajectory (see Fig.~\ref{fig:eq_1contours}), the new area can be approximated
by integrating Eq.~\eqref{eq:a_oint} along the upper leg of the separatrix.
Pictorially, this can be seen in the bottom panels of Fig.~\ref{fig:ad_21}.

\subsubsection{Governing Principle: Probabilistic Separatrix Crossing}

When a trajectory experiences separatrix crossing, it transitions into nearby
zones probabilistically. This process is studied in the
adiabatic limit by \citet{henrard1982} and \citet{henrard1987}. Their results
may be summarized as follows: if zone $i$ is shrinking while adjacent zones $j,
k$ are expanding such that the sum of their areas is constant, the probabilities
of transition from zone $i$ to zones $j$ and $k$ are given by
\begin{subequations}\label{eq:henrard_hop}
    \begin{align}
        \Pr\p{i \to j} = -\frac{\pdil{\mathcal{A}_{\rm
            j}}{\eta}}{\pdil{\mathcal{A}_{\rm i}}{\eta}}, \\
        \Pr\p{i \to k} = -\frac{\pdil{\mathcal{A}_{\rm
            k}}{\eta}}{\pdil{\mathcal{A}_{\rm i}}{\eta}}.
    \end{align}
\end{subequations}
Note that $\Pr \p{i \to j} + \Pr\p{i \to k} = 1$.
Eqs.~\eqref{eq:henrard_hop} can be used in conjunction with
Eqs.~\eqref{se:area_ward} to understand for what initial conditions each track
can be observed and with what probabilities.

As an example, consider a system in zone II in panel (d) of
Fig.~\ref{fig:eq_1contours}. As $\eta$ decreases, zone II will shrink while
zones I and III will expand until the trajectory crosses the separatrix. Suppose
the trajectory exits zone II at some $\eta_\star$, then the probability of the
II $\to$ I transition is $\Pr\p{\rm II \to I} = -\dot{A}_{\rm I} / \dot{A}_{\rm
II}$, while the II $\to$ III transition occurs with probability $\Pr\p{\rm II
\to III} = -\dot{A}_{\rm III} / \dot{A}_{\rm II}$.

\subsubsection{Evolutionary Trajectories}\label{sss:evol_traj}

\begin{figure}
    \centering
    \includegraphics[width=0.47\textwidth]{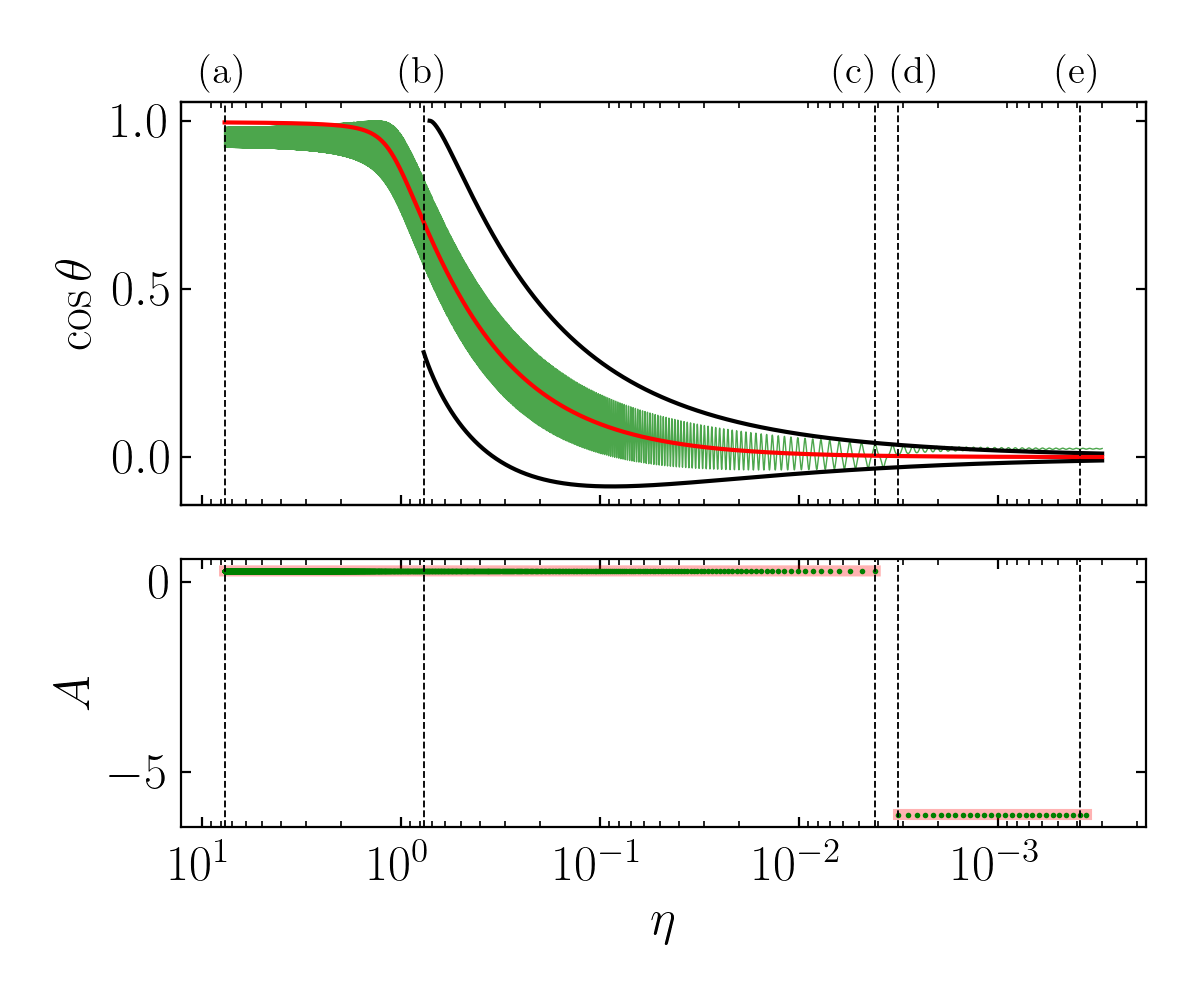}

    \includegraphics[width=0.47\textwidth]{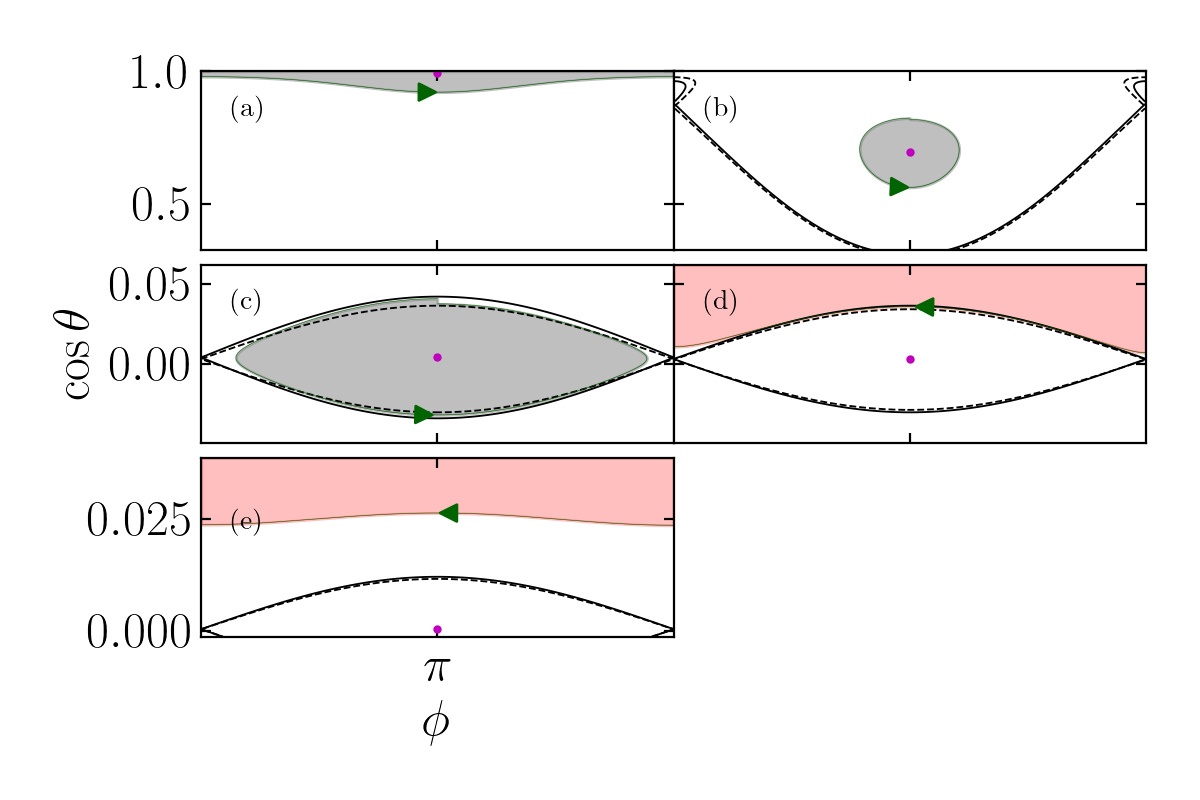}
    \caption{An example of the II $\to$ I evolutionary track for $I = 5^\circ$
    and $\theta_{\rm sd, i} = 17.2^\circ$. Upper panel: The thin green line shows
    $\cos \theta$ as a function of $\eta$, obtained by numerical integration
    (with $\epsilon = 3 \times 10^{-4}$). Overlaid are the location of Cassini
    State 2 (dashed red) and the upper and lower bounds on the separatrix
    (dotted black). The trajectory tracks CS2 to a final obliquity of
    $88.57^\circ$. The black vertical dashed lines denote instants in the
    simulation  portrayed in bottom panels. Middle panel: The enclosed
    separatrix area obtained by integrating the simulated trajectory (green
    dots) and adiabatic theory (red line). Lower plot: Snapshots in $\p{\cos
    \theta, \phi}$ phase space of one circulation/libration cycle of the
    trajectory, shown in dark green with an arrow indicating direction. The
    snapshots correspond to the start of the simulation (a), the appearance of
    the separatrix (b), two panels depicting the separatrix crossing process
    (c-d), and a final snapshot at $\eta = 10^{-3.5}$ (e). The separatrices at
    the beginning and end of the portrayed cycle in each snapshot are shown in
    solid/dashed black lines respectively. Also labeled is CS2 at the start of
    each cycle (filled red circle). Finally, the enclosed phase space area is
    shaded in grey ($A > 0$) and red ($A < 0$).}\label{fig:ad_21}
\end{figure}

\begin{figure}
    \centering
    \includegraphics[width=0.47\textwidth]{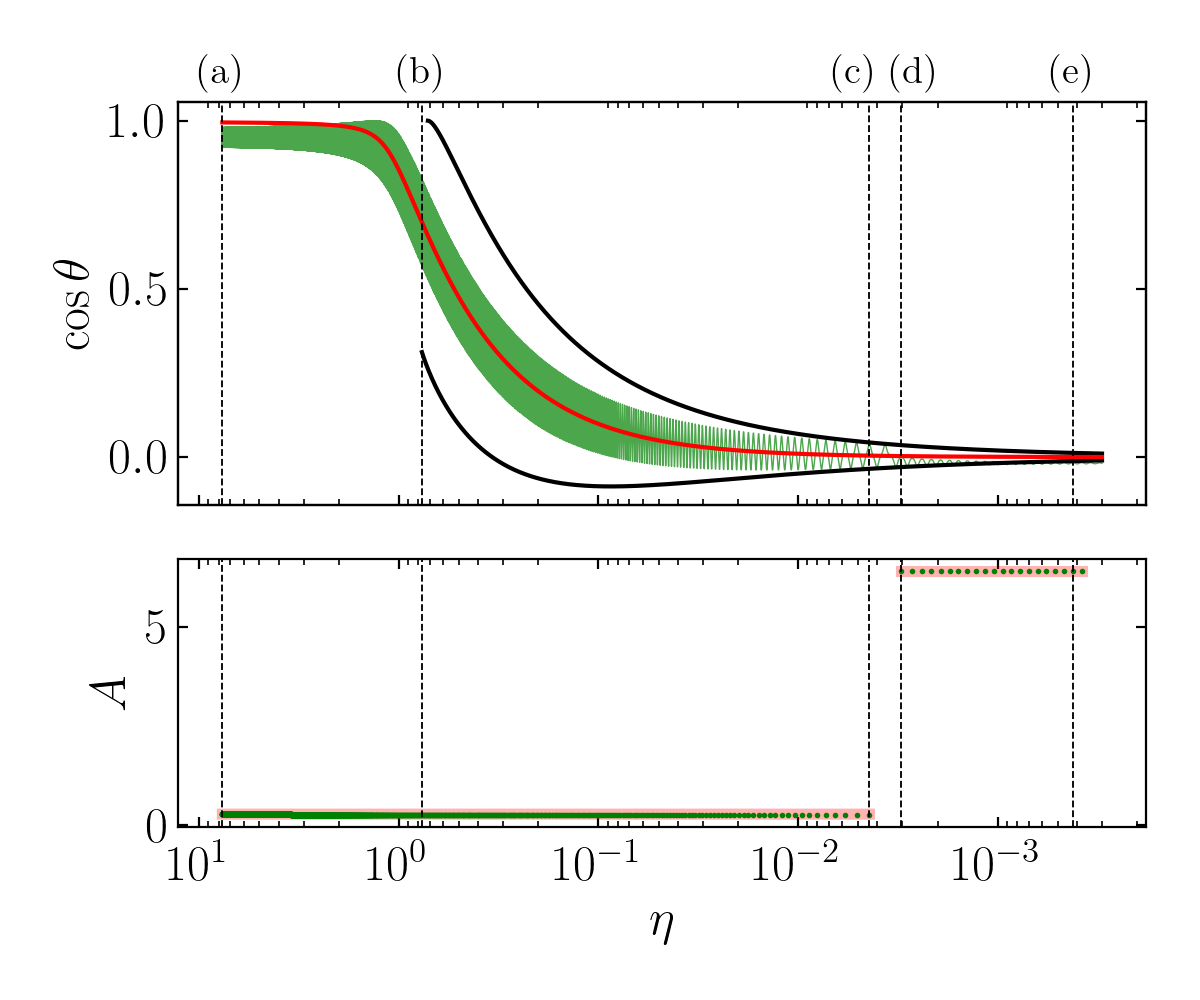}

    \includegraphics[width=0.47\textwidth]{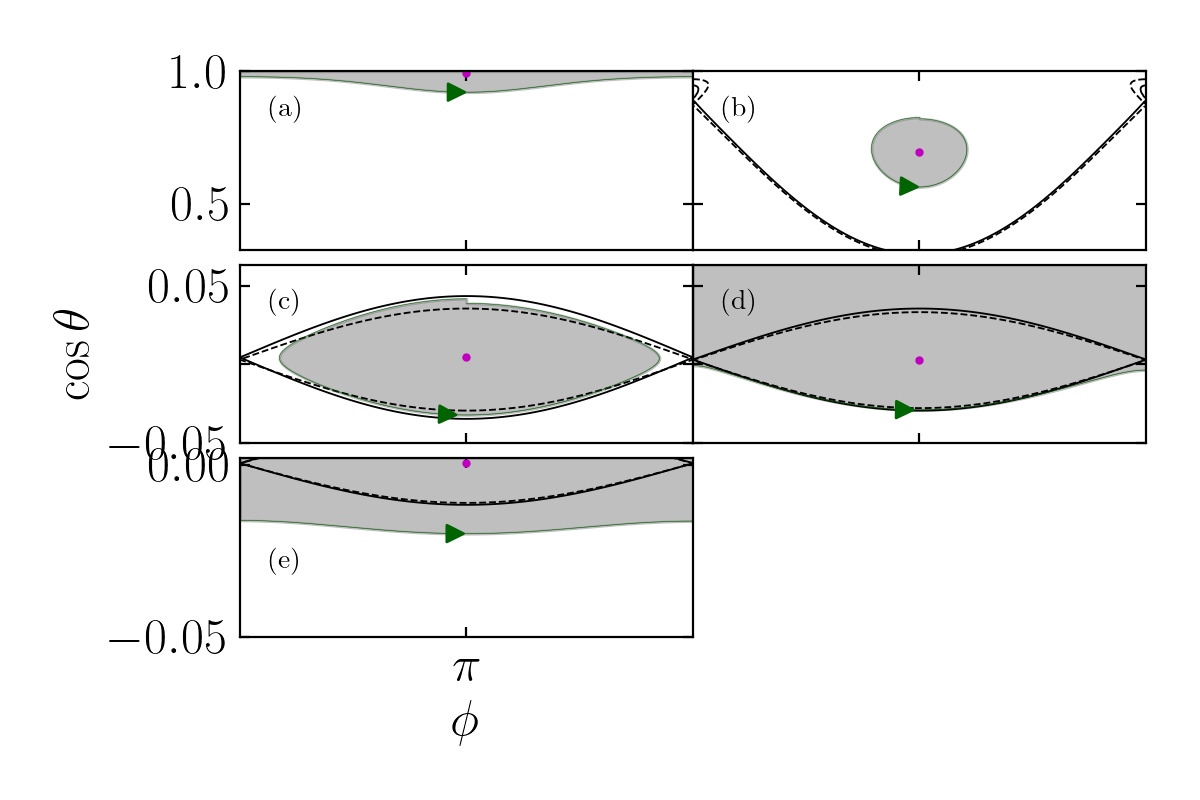}
    \caption{Same as Fig.~\ref{fig:ad_21} but for the II $\to$ III track.
    $\theta_{\rm sd, i} = 17.2^\circ$ and $\epsilon = 3.01 \times
    10^{-4}$.}\label{fig:ad_23}
\end{figure}

\begin{figure}
    \centering
    \includegraphics[width=0.47\textwidth]{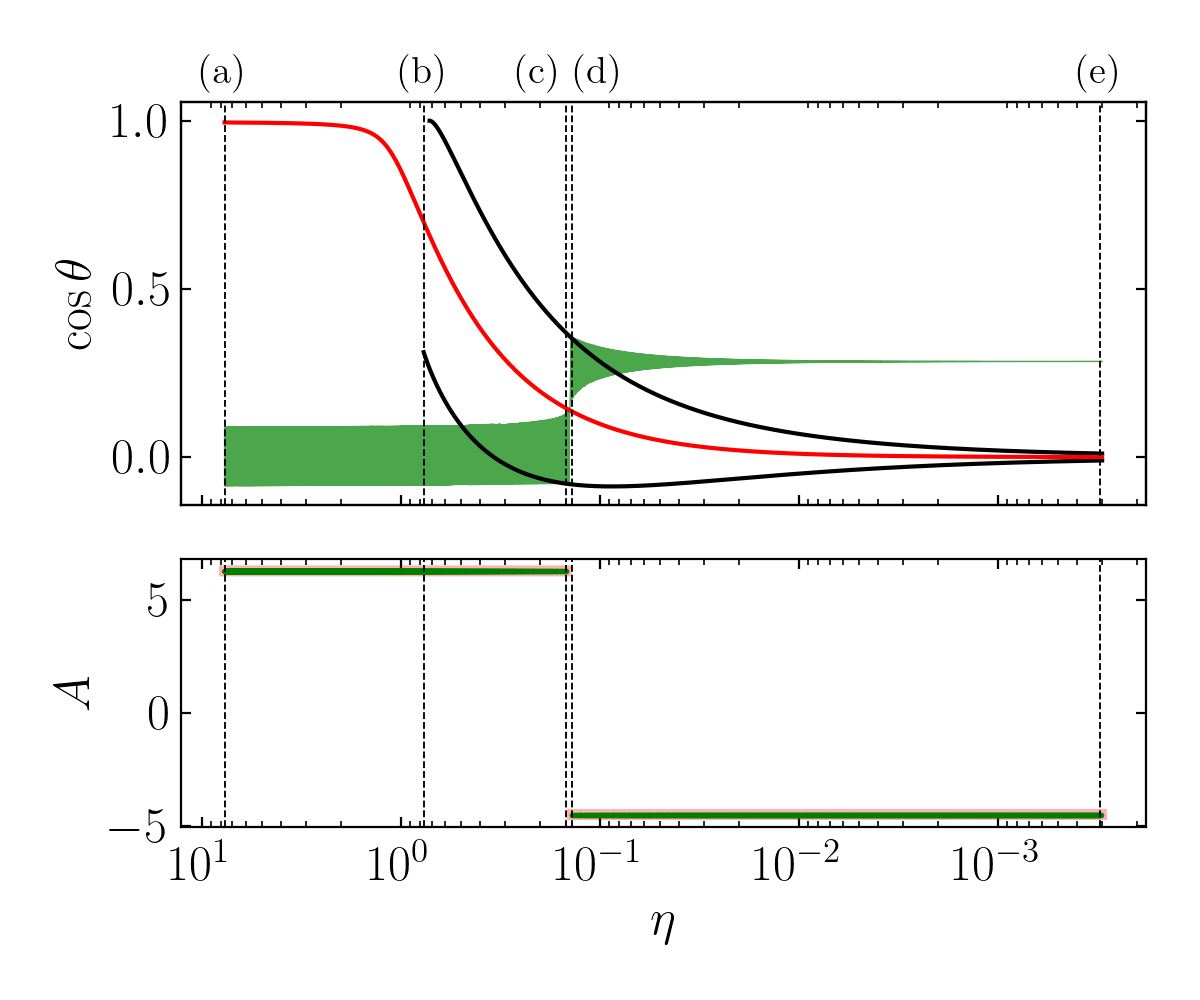}

    \includegraphics[width=0.47\textwidth]{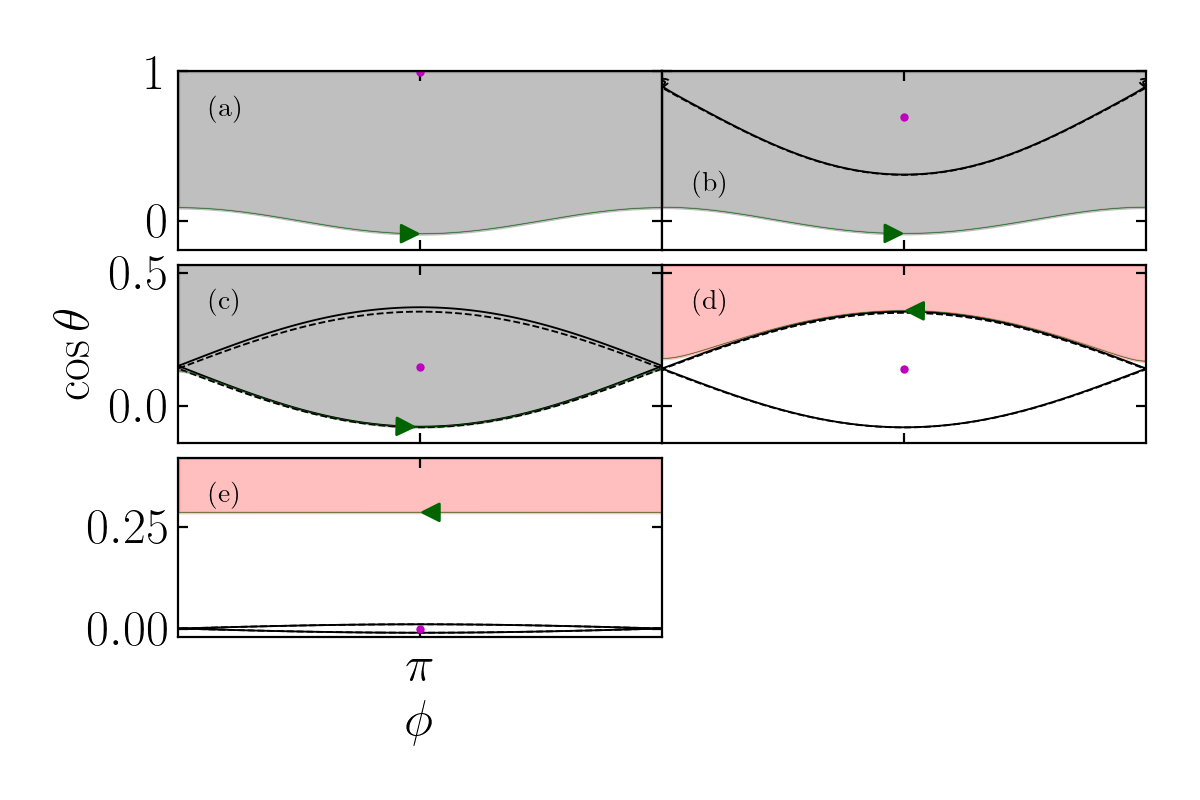}
    \caption{Same as Fig.~\ref{fig:ad_21} but for the III $\to$ I track.
    $\theta_{\rm sd, i} = 89.1^\circ$, and $\epsilon = 3 \times
    10^{-4}$.}\label{fig:ad_31}
\end{figure}

\begin{figure}
    \centering
    \includegraphics[width=0.47\textwidth]{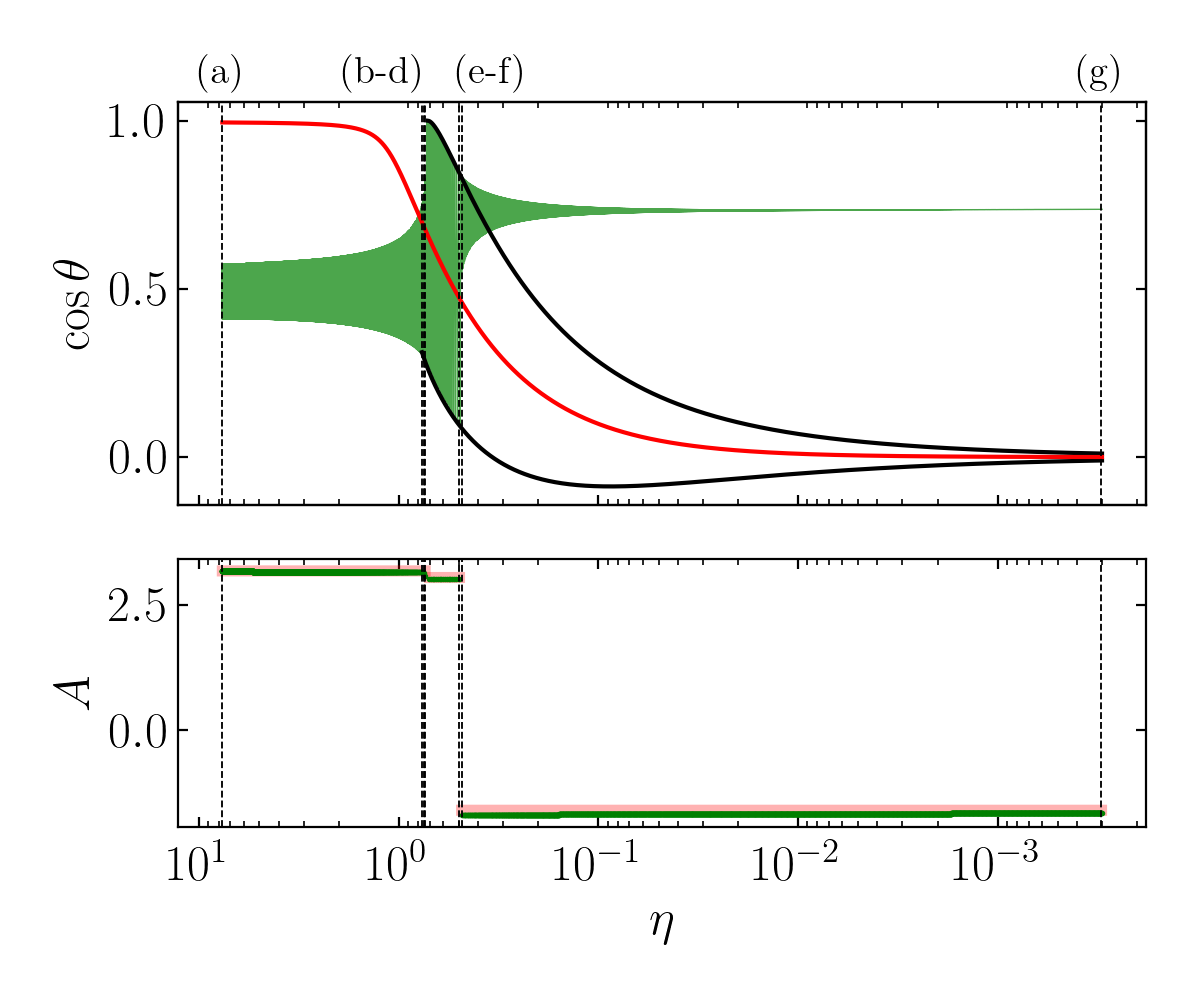}

    \includegraphics[width=0.47\textwidth]{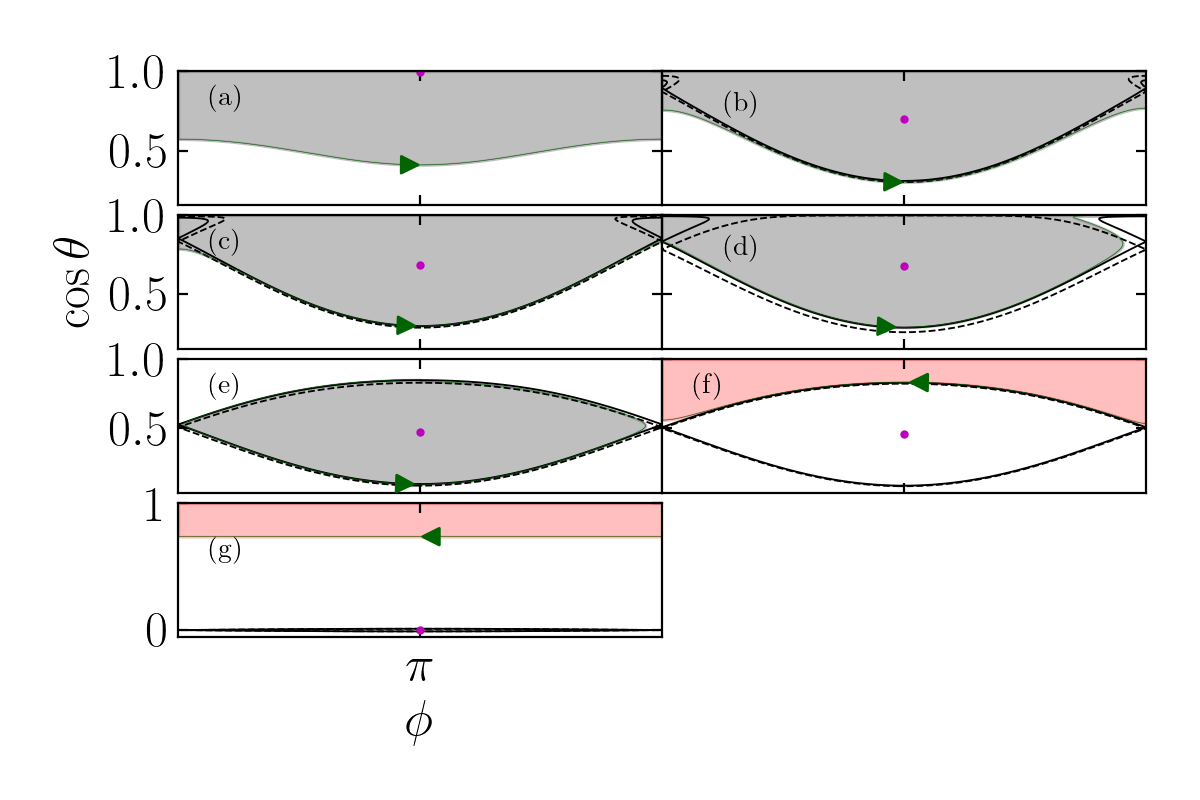}
    \caption{Same as Fig.~\ref{fig:ad_21} but for the III $\to$ II $\to$ I
    track. $\theta_{\rm sd, i} = 60^\circ$, and $\epsilon = 3.14 \times
    10^{-4}$. Two separatrix crossings are shown, in panels (c-d) and
    (e-f).}\label{fig:ad_321}
\end{figure}

Returning to the evolution of $\uv{s}$, we can classify trajectories by the
sequence of separatrix encounters. Initially, in the $\eta > \eta_{\rm c}$
regime, only zones II and III exist; as $\eta \to 0$, only zones I and III exist
(see Fig.~\ref{fig:eq_1contours}). There are five distinct evolutionary tracks:
\begin{enumerate}
    \item II $\to$ I (see Fig.~\ref{fig:ad_21} for an example). The spin axis
        $\uv{s}$ initially circulates in zone II (snapshot a), and then starts
        librating about CS2 as $\eta$ decreases (snapshot b), enclosing some
        initial phase space area $A_{\rm i}$. This libration continues until the
        separatrix expands (due to decreasing $\eta$) to ``touch'' the
        trajectory (snapshot c), at which $\mathcal{A}_{\rm II}(\eta_\star) =
        A_{\rm i}$. As $\hat{s}$ moves to a circulating trajectory in zone I
        immediately bordering the separatrix, it will encompass
        $-\mathcal{A}_{\rm I}(\eta_\star)$ phase space area. The final obliquity
        $\theta_{\rm f}$ is then given by Eq.~\eqref{eq:qfaf}, with $A_{\rm f} =
        -\mathcal{A}_{\rm I}\p{\eta_\star}$. An analytical approximation to
        $\theta_{\rm f}$ is derived in Appendix~\ref{s:ad_approx} and is
        \begin{equation}
            \p{\cos \theta_{\rm f}}_{\rm II \to I} \simeq
                \p{\frac{\pi \theta_{\rm sd, i}^2}{16}}^2 \cot I
                    + \frac{\theta_{\rm sd, i}^2}{4}.\label{eq:qf_21_approx}
        \end{equation}
        The transition probability is
        \begin{equation}
            \Pr\p{\mathrm{II} \to \mathrm{I}} = -\p{
                \frac{\pdil{\mathcal{A}_{\rm I}}{\eta}}{\pdil{\mathcal{A}_{\rm
                    II}}{\eta}}} _{\eta = \eta_{\star}}.
        \end{equation}
        This track can only occur when the initial condition begins in zone II,
        requiring $A_{\rm i} < \mathcal{A}_{\rm II}(\eta_{\rm c})$, where
        $\mathcal{A}_{\rm II}\p{\eta_c}$ is given by Eq.~\eqref{eq:A2} evaluated
        at $\eta = \eta_c$. Since $\pdil{\mathcal{A}_{\rm I}}{\eta} < 0$
        everywhere, while $\pdil{\mathcal{A}_{\rm II}}{\eta} > 0$ at all
        possible $\eta_\star$ for an initial condition starting in zone II, this
        track always has nonzero probability.

    \item II $\to$ III (see Fig.~\ref{fig:ad_23}). This track is similar to the
        II $\to$ I track; the only difference is that, upon separatrix
        encounter, the trajectory follows the circulating trajectory in zone III
        bordering the separatrix, upon which it will encompass area
        $\mathcal{A}_{\rm I}(\eta_\star) + \mathcal{A}_{\rm II}(\eta_\star) =
        A_{\rm f}$. The final obliquity is still given by Eq.~\eqref{eq:qfaf},
        and the analytical approximation derived in Appendix~\ref{s:ad_approx}
        is
        \begin{equation}
            \p{\cos \theta_{\rm f}}_{\rm II \to III} \simeq
                \p{\frac{\pi \theta_{\rm sd, i}^2}{16}}^2 \cot I
                    - \frac{\theta_{\rm sd, i}^2}{4}.\label{eq:qf_23_approx}
        \end{equation}
        The transition probability is
        \begin{equation}
            \Pr\p{\mathrm{II} \to \mathrm{III}} = -\p{
                \frac{\pdil{\mathcal{A}_{\rm III}}{\eta}}{\pdil{\mathcal{A}_{\rm
                    II}}{\eta}}} _{\eta = \eta_{\star}}.
        \end{equation}
        Again, this track can only occur when $A_{\rm i} < \mathcal{A}_{\rm
        II}(\eta_{\rm c})$, but a further constraint arises when we consider the
        transition probability. Upon examination of Fig.~\ref{fig:eq_areas}, it
        is clear that $\pdil{\mathcal{A}_{\rm III}}{\eta} > 0$ for a large range
        of $\eta$, which would give a negative transition probability---implying
        a forbidden transition. Define
        \begin{equation}
            \eta_{\min, III} \equiv \argmin \mathcal{A}_{\rm III}(\eta)
                \label{eq:eta_minIII},
        \end{equation}
        which is labeled in Fig.~\ref{fig:eq_areas}. Thus, the II $\to$ III
        track is permitted only if $\eta_\star <, \eta_{\rm \min, III}$.

    \item III $\to$ I (see Fig.~\ref{fig:ad_31}). The trajectory encounters the
        separatrix when $\mathcal{A}_{\rm I}(\eta_\star) + \mathcal{A}_{\rm
        II}(\eta_\star) = A_{\rm i}$, upon which it transitions to a zone I
        trajectory enclosing $A_{\rm f} = -\mathcal{A}_{\rm I}$. The final
        obliquity is again given by Eq.~\eqref{eq:qfaf}.

        This track can only occur if $A_{\rm i} > \mathcal{A}_{\rm II}(\eta_{\rm
        c})$, but is also constrained by requiring $A_{\rm i}$ be sufficiently
        small so that it will encounter the separatrix (if $A_{\rm i}$ is too
        large, it will never encounter the separatrix, and we simply have a III
        $\to$ III transition). This condition is $A_{\rm i} < \max
        \p{\mathcal{A}_{\rm I} + \mathcal{A}_{\rm II}} = 4\pi - \min
        \p{\mathcal{A}_{\rm III}}$. Since $\pdil{\mathcal{A}_{\rm I}}{\eta} < 0$
        and $\pdil{\mathcal{A}_{\rm III}}{\eta} > 0$ for all accessible
        $\eta_{\star}$, this track is always permitted.

    \item III $\to$ II $\to$ I (see Fig.~\ref{fig:ad_321}). That
        $\mathcal{A}_{\rm II}(\eta)$ is not a monotonic function of $\eta$ (see
        Fig.~\ref{fig:eq_areas}) is key to the existence of this track. Consider
        a trajectory originating in zone III that first encounters the
        separatrix at $\eta_1$, when $\mathcal{A}_{\rm I}(\eta_1) +
        \mathcal{A}_{\rm II}(\eta_1) = A_{\rm i}$, such that it transitions into
        zone II enclosing intermediate phase space area $A_{\rm m} =
        \mathcal{A}_{\rm II}(\eta_1)$. Such a transition has probability
        \begin{equation}
            \Pr\p{\rm III \to II} = -\p{
                \frac{\pdil{\mathcal{A}_{\rm II}}{\eta}}{\pdil{\mathcal{A}_{\rm
                    III}}{\eta}}} _{\eta = \eta_1},
        \end{equation}
        which is nonnegative (i.e.\ the transition is permitted) if $\eta_1 \in
        [\eta_{\rm \max, II}, \eta_{\rm c}]$. Equivalently, this requires
        $A_{\rm i} \in \s{\mathcal{A}_{\rm II}\p{\eta_{\rm c}} ,
        \mathcal{A}_{\rm II, \max}}$. Then, as $\eta$ continues to decrease, a
        second $\eta_2$ value exists for which $A_{\rm m} = \mathcal{A}_{\rm
        II}(\eta_2)$, upon which the trajectory is ejected to zone I and $A_{\rm
        f} = -\mathcal{A}_{\rm I}(\eta_2)$. Note that $\eta_2 < \eta_{\rm \max,
        II}$ necessarily, as zone II must be shrinking in order for the
        trajectory to be ejected. The final obliquity is given by
        Eq.~\eqref{eq:qfaf}. Graphical inspection of Fig.~\ref{fig:eq_areas}
        shows that $\pdil{\mathcal{A}_{\rm II}}{\eta}$ and
        $\pdil{\mathcal{A}_{\rm III}}{\eta}$ have the same signs for $\eta <
        \eta_{\rm \max, II}$, and therefore the complementary II $\to$ I
        transition is guaranteed. Overall, the III $\to$ II $\to$ I track is
        permitted so long as the first transition is permitted, or $A_{\rm i}
        \in \s{\mathcal{A}_{\rm II}\p{\eta_{\rm c}} , \mathcal{A}_{\rm II,
        \max}}$.

    \item III $\to$ III\@. This track is the trivial case where no separatrix
        encounter occurs, and $A$ is constant throughout the evolution ($A_{\rm
        f} = A_{\rm i}$) except for a jump by $4\pi$ when crossing the South
        pole ($\cos \theta = -1$) due to the coordinate singularity. This
        requires $A_{\rm i} > \max \p{\mathcal{A}_{\rm I} + \mathcal{A}_{\rm
        II}}$. In the limit of $\eta_{\rm i} \to \infty$ and $ \eta_{\rm f} \to
        0$ we have $\theta_{\rm f} = \theta_{\rm sd, i}$. For finite $\eta_{\rm
        i}$, the initial enclosed phase space area for III $\to$ III
        trajectories is not given exactly by using $\theta = \theta_{\rm sd, i}$
        in Eq.~\eqref{eq:ai_qsd}. This is because the initial orbits for such
        trajectories are better described as librating about CS3 with angle of
        libration $\Delta \theta - \theta_{\rm sd, i}$ rather than about CS2
        with angle of libration $\theta_{\rm sd, i}$. Here, $\Delta \theta$ is
        the angular distance between CS2 and CS3 and is not equal to $180^\circ$
        except when $\eta_{\rm i} \to \infty$. This finite-$\eta_{\rm i}$ effect
        is responsible for the small cusp at the very right ($\theta_{\rm sd, i}
        \to 180^\circ$) of Figs.~\ref{fig:ad_ensemble}
        and~\ref{fig:3_ensemble_20_35}.
\end{enumerate}

In summary, starting from an initial condition with phase space area $A_{\rm i}$
at $\eta = \eta_{\rm i} \gg 1$, the five evolutionary tracks are:
\begin{enumerate}
    \item $A_{\rm i} \in \s{0, \mathcal{A}_{\rm II}\p{\eta_{\rm \min, III}}}$:
        Both the II $\to$ III and the II $\to$ I tracks are possible.

    \item $A_{\rm i} \in \s{\mathcal{A}_{\rm II}\p{\eta_{\rm \min, III}},
        \mathcal{A}_{\rm II}(\eta_{\rm c})}$: Only the II $\to$ I track.

    \item $A_{\rm i} \in \s{\mathcal{A}_{\rm II}(\eta_{\rm c}), \mathcal{A}_{\rm
        II, \max}}$: Both the III $\to$ I and III $\to$ II $\to$ I are possible.

    \item $A_{\rm i} \in \s{\mathcal{A}_{\rm II, \max}, \max \p{\mathcal{A}_{\rm
        I} + \mathcal{A}_{\rm II}}}$: Only the III $\to$ I track.

    \item $A_{\rm i} > \max \p{\mathcal{A}_{\rm I} + \mathcal{A}_{\rm II}}$:
        Only the III $\to$ III track.
\end{enumerate}
In all cases, the corresponding ranges for $\theta_{\rm sd,i}$ can be computed via
Eq.~\eqref{eq:ai_qsd}. The boundaries between these ranges are overplotted in
Fig.~\ref{fig:ad_ensemble}, where they can be seen to agree well with the
numerical results.

\section{Nonadiabatic Effects}\label{s:nonad}

In Section~\ref{s:ad}, we have examined the spin axis evolution in the limit
where $\epsilon \ll \epsilon_{\rm c}$ [see Eq.~\eqref{eq:ad_constr}] and the
evolution is mostly adiabatic (except at separatrix crossings). We now consider
nonadiabatic effects.

\subsection{Transition to Non-adiabaticity: Results for $\epsilon\lesssim
\epsilon_{\rm c}$}\label{ss:transition}

\begin{figure}
    \centering
    \includegraphics[width=0.47\textwidth]{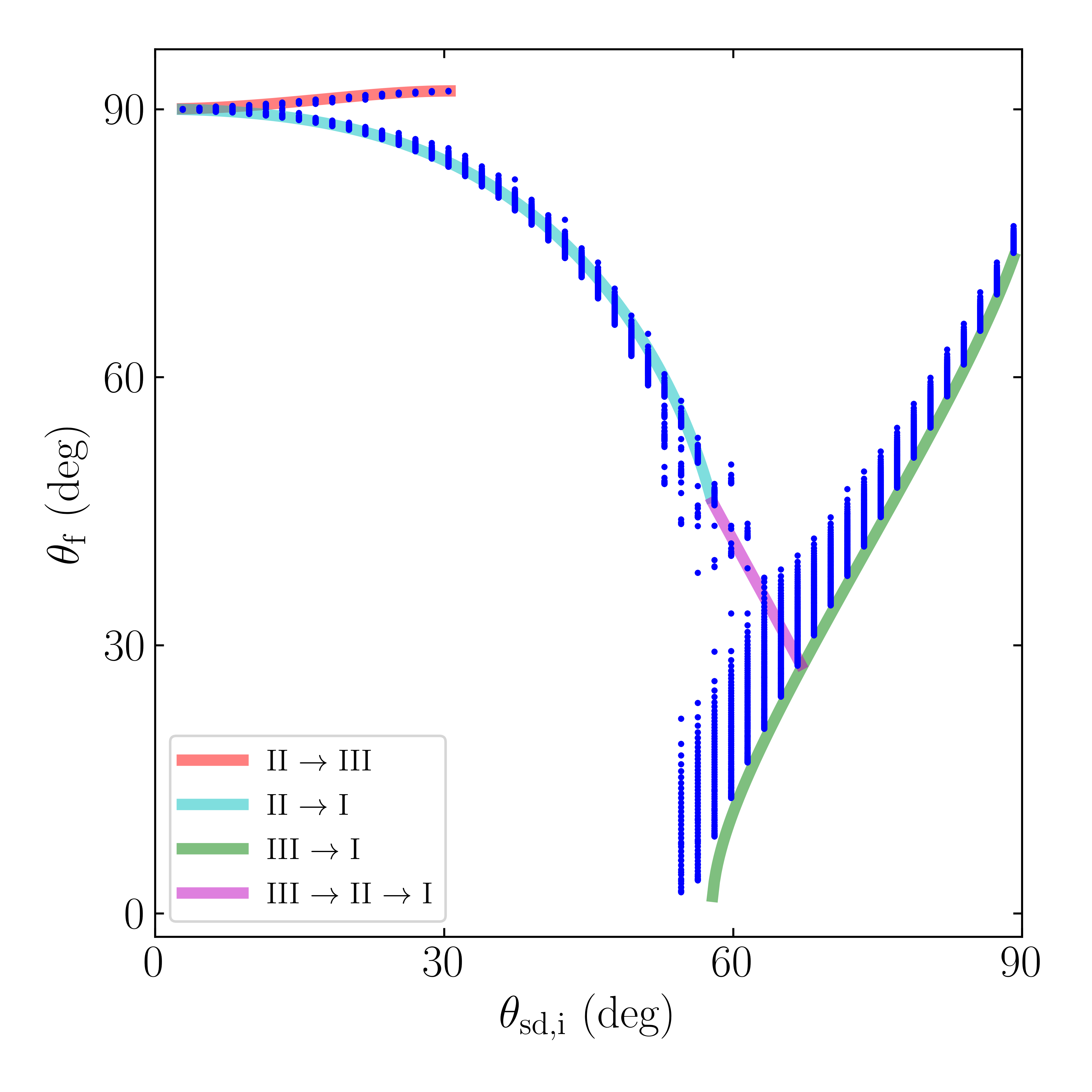}
    \caption{Same as Fig.~\ref{fig:ad_ensemble} but for $\epsilon = 10^{-2.5}$
    and restricting $\theta_{\rm sd, i} < 90^\circ$ (blue dots). The colored
    solid lines are analytical adiabatic results (same as shown in
    Fig.~\ref{fig:ad_ensemble}). A larger spread from the adiabatic tracks is
    observed in the numerical results due to the non-adiabaticity effect.
    }\label{fig:3_ensemble_05_25}
\end{figure}

\begin{figure}
    \centering
    \includegraphics[width=0.47\textwidth]{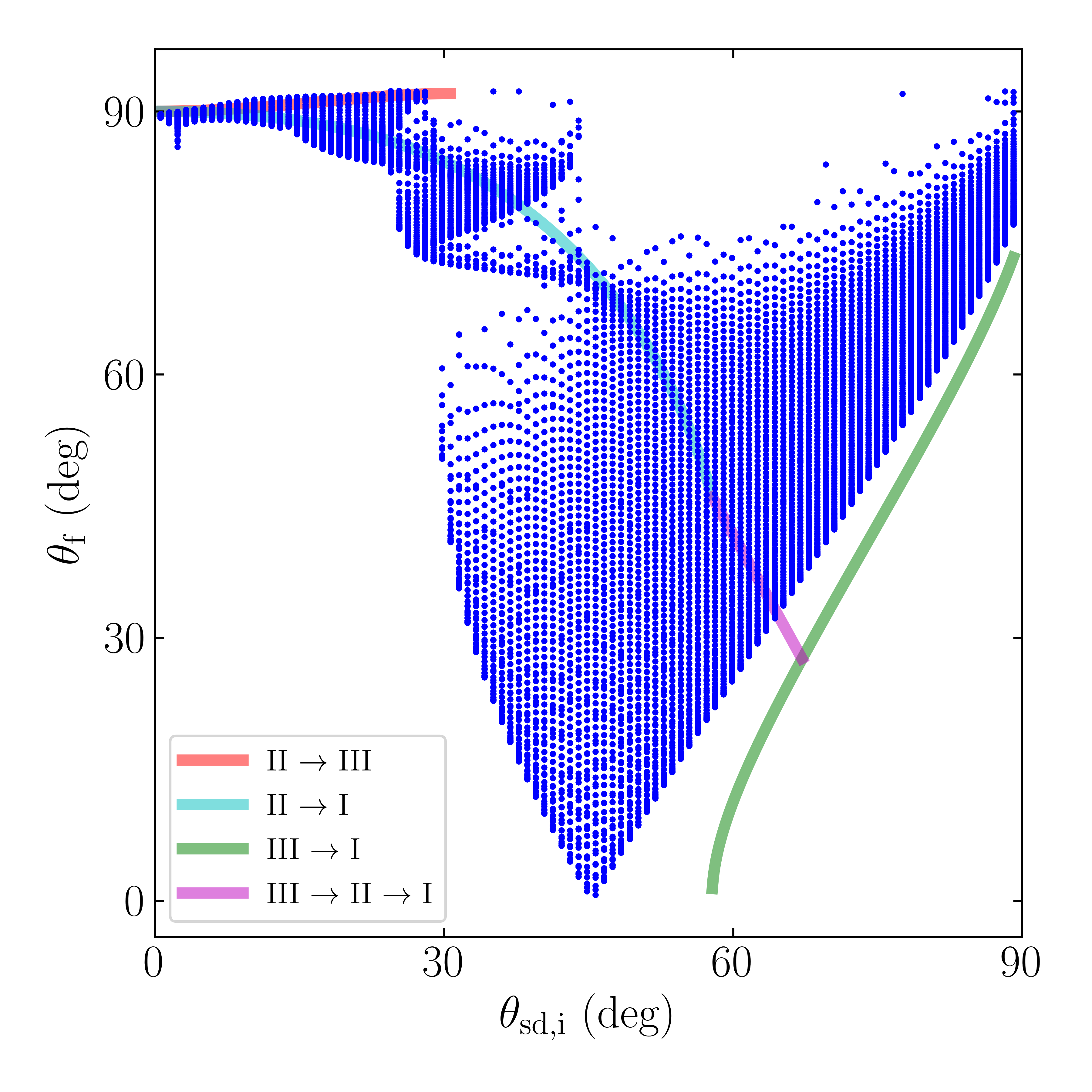}
    \caption{Same as Fig.~\ref{fig:3_ensemble_05_25} but for $\epsilon =
    10^{-1.5}$ (i.e.\ larger non-adiabaticity effect). Some small resemblance to
    the adiabatic tracks remains, and the deviations appear to have a banded
    structure.}\label{fig:3_ensemble_05_15}
\end{figure}

To illustrate the transition to nonadiabaticity, we carried out a suite of
numerical calculations for several values of $\epsilon$. The results for two of
these values are shown in Figs.~\ref{fig:3_ensemble_05_25}
and~\ref{fig:3_ensemble_05_15}.

As $\epsilon$ increases (see Fig.~\ref{fig:3_ensemble_05_25}), nonadiabaticity
manifests as a larger scatter of final obliquities near the tracks predicted
from adiabatic evolution. This scatter first sets in for trajectories starting
in zone III, as these trajectories encounter the separatrix at larger $\eta$
compared to those originating in zone II\@. This means the obliquity of CS2
$\theta_2$ is smaller for these trajectories, and the adiabaticity criterion is
stricter [see Eq.~\eqref{eq:ad_constr}]. Physically, approaching the
adiabaticity criterion corresponds to the separatrix crossing process becoming
sensitive to the \emph{phase} of the libration/circulation cycle at the
crossing: if the trajectory crosses the separatrix when the obliquity is at its
maximum, the final obliquity will also be relatively larger.

As $\epsilon$ increases further (see Fig.~\ref{fig:3_ensemble_05_15}) but still
marginally satisfies the weak adiabaticity criterion [Eq.~\eqref{eq:ad_constr}],
the scatter in $\theta_{\rm f}$ continues to widen. The horizontal banded
structure of the final obliquities is a consequence of even stronger phase
sensitivity during separatrix crossing: trajectories cross the separatrix at
similar phases evolve to similar final obliquities that only depend weakly on
on $\theta_{\rm sd, i}$. Finally, in
Fig.~\ref{fig:3_ensemble_05_15}, the bottom edge of the data and the III $\to$ I
track deviate very noticeably. This non-adiabatic effect is the result of the
separatrix evolving significantly within the separatrix-crossing orbit, as the
tracks computed in Section~\ref{s:ad} assume that $\eta$ is constant throughout
the separatrix-crossing orbit.

\begin{figure}
    \centering
    \includegraphics[width=0.47\textwidth]{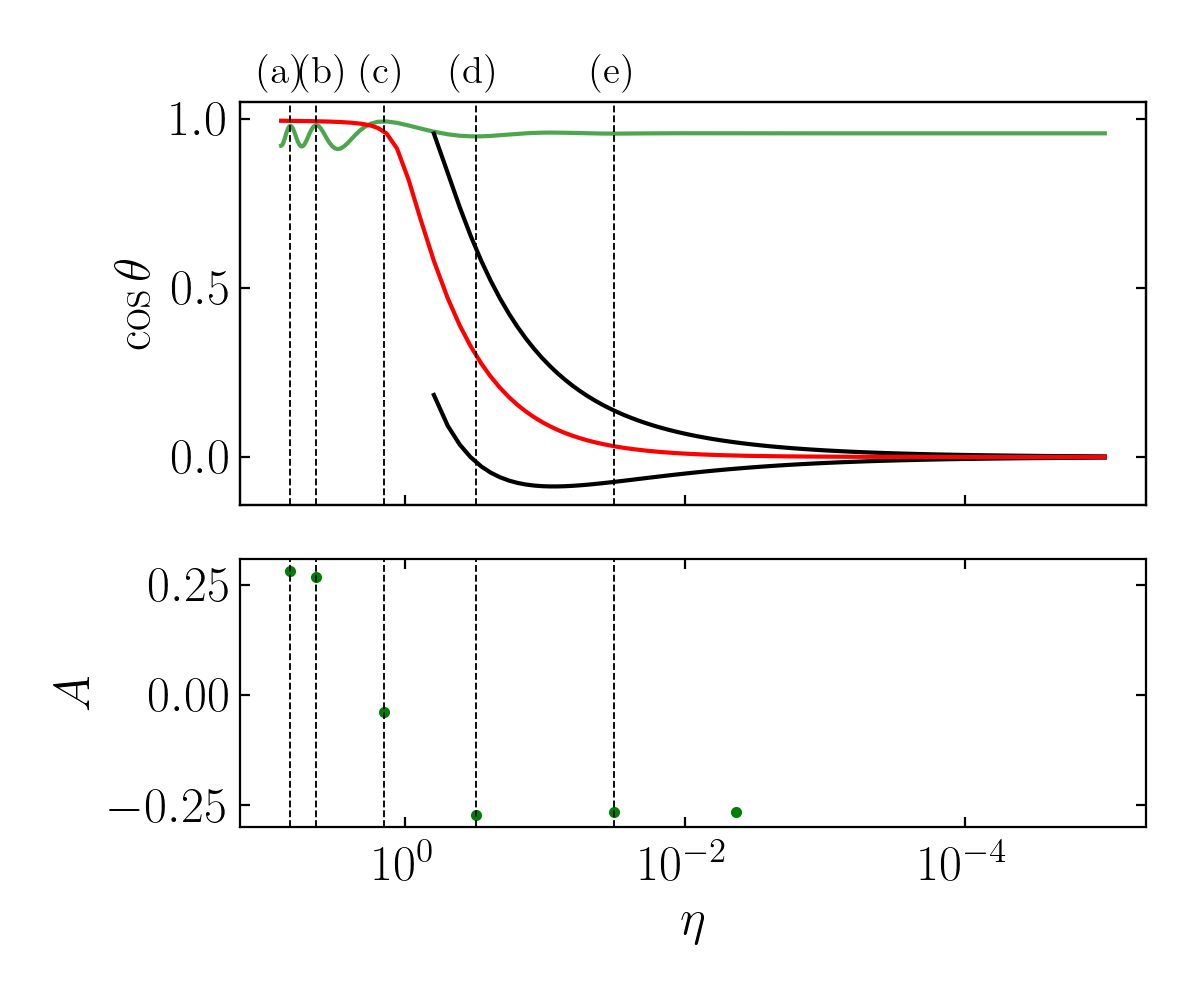}

    \includegraphics[width=0.47\textwidth]{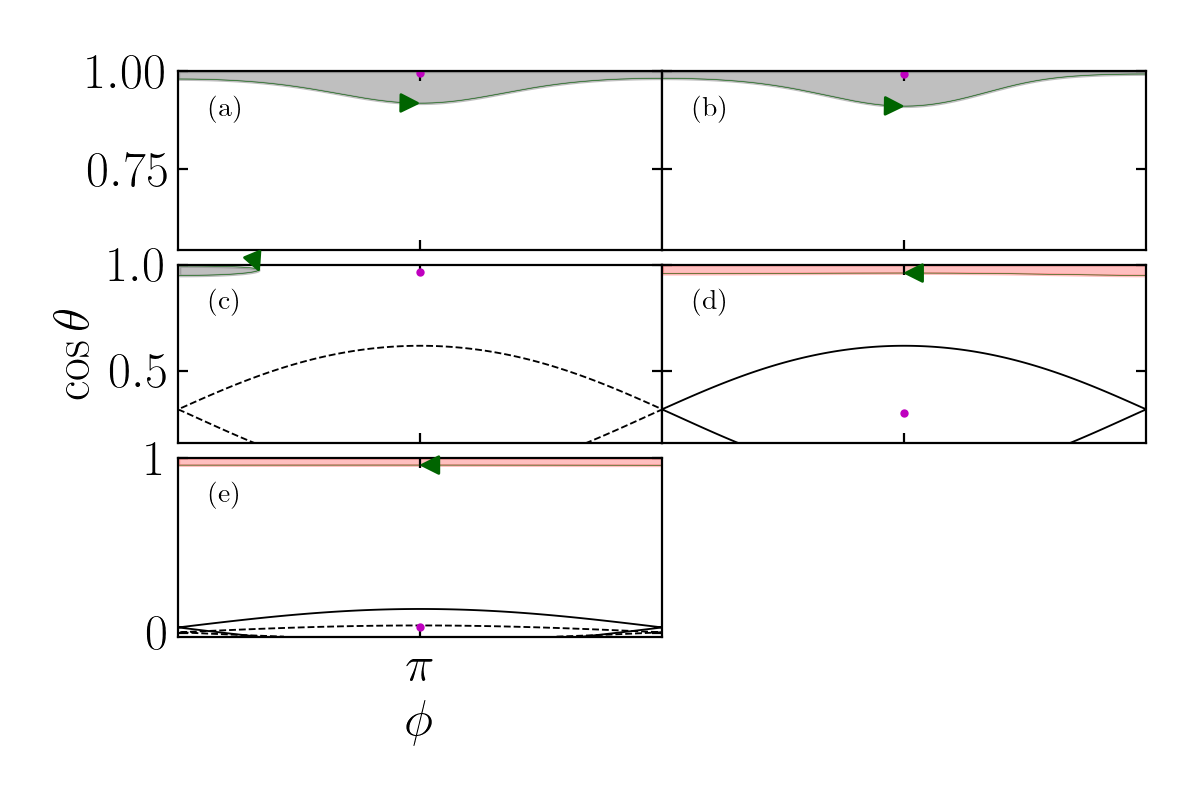}
    \caption{Same as Fig.~\ref{fig:ad_21} but for a nonadiabatic case, with $\epsilon =
    0.3$. In the top panel, it is evident that the libration cycle about CS2 is
    unable to keep up with the swift migration of CS2 as $\eta$ changes,
    decreasing the obliquity excitation compared to the adiabatic simulation. In
    the middle panel, the trajectory only undergoes six libration/circulation
    cycles before $\eta < 10^{-5}$, and the enclosed phase space area is
    clearly not conserved. In the bottom panel, we can see that individual
    trajectories do not lie along level curves of the Hamiltonian, as the
    Hamiltonian phase space changes quickly compared to the period of
    circulation cycles.}\label{fig:nonad_traj}
\end{figure}

A sample trajectory following in the style of Fig.~\ref{fig:ad_21} but for
$\epsilon = 0.3$ (violating even weak adiabaticity) is provided in
Fig.~\ref{fig:nonad_traj}. It is clear that the trajectory does not track the
level curves of the Hamiltonian during each individual snapshot. This results
from CS2 migrating more quickly than the trajectory can librate about CS2,
violating the weak adiabaticity criterion.

\subsection{Non-adiabatic Evolution:
Result for $\epsilon\gtrsim \epsilon_{\rm c}$}

In general, numerical calculations are needed to determine the non-adiabatic
obliquity evolution ($\epsilon\gtrsim \epsilon_{\rm c}$). However, some
analytical results can still be obtained when the obliquity change is small.

\begin{figure}
    \centering
    \includegraphics[width=0.47\textwidth]{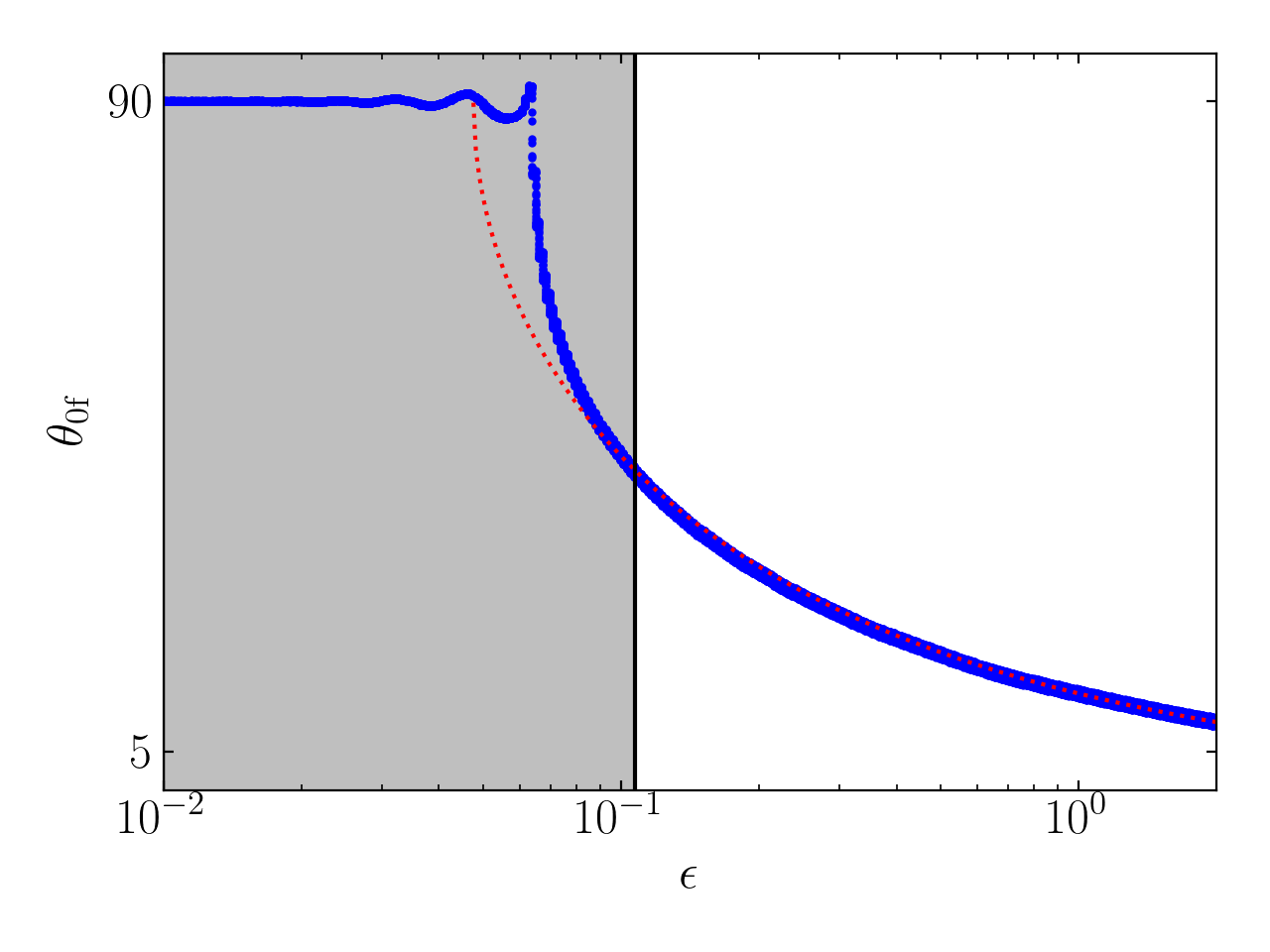}
    \caption{Final obliquity $\theta_{\rm  f}$ as a function of $\epsilon$ for
    $\theta_{\rm sd,i} = 0$ and $I = 5^\circ$. The shaded area, bordered by the
    black line, corresponds to the adiabatic regime estimated by
    Eq.~\eqref{eq:ad_constr}. The blue dots are numerical results, and the red
    dashed line corresponds to Eq.~\eqref{eq:theta_2f}, which is in good
    agreement with numerical results for $\epsilon > \epsilon_{\rm c} \approx
    0.1$ (the nonadiabatic regime). Note that $\theta_{\rm f} \simeq 90^\circ$
    in the adiabatic regime ($\epsilon\ll \epsilon_{\rm
    c}$).}\label{fig:nonad_3_scan}
\end{figure}

We start from Eq.~\eqref{eq:dsdt_base}, which governs the evolution of the spin
axis in the rotating frame. We choose coordinate axes such that $\uv{l} =
\uv{z}$ and $ \uv{l}_{\rm d} = \uv{z} \cos I + \uv{x}\sin I$, giving
\begin{equation}
    \p{\rd{\uv{s}}{\tau}}_{\rm rot} = \s{
        \p{\eta \cos I - \cos \theta}\uv{z}
            + \eta \sin I \,\uv{x}} \times \uv{s}.
\end{equation}
Defining $S = {\hat s}_{\rm x} + i{\hat s}_{\rm y}$, we find
\begin{equation}
    \rd{S}{\tau} = i\p{\eta\cos I - \cos \theta}S
        - i \eta \sin I\cos\theta.\label{eq:nonad_ode}
\end{equation}
To proceed, we assume the obliquity is roughly constant, $\cos \theta \approx
\cos \theta_{\rm i}$. Eq.~\eqref{eq:nonad_ode} can then be solved explicitly,
starting from the initial value $S(\tau_{\rm i})$:
\begin{equation}
    S(\tau)e^{-i\Phi(\tau)} - S(\tau_{\rm i})
        \simeq -i\sin I \cos \theta_{\rm i}
            \int_{\tau_{\rm i}}^\tau \eta(\tau')
            e^{-i\Phi(\tau')}\;\mathrm{d}\tau',
\end{equation}
where
\begin{equation}
    \Phi(\tau) \equiv \int_{\tau_{\rm i}}^\tau \p{\eta(\tau') \cos I
        - \cos \theta_{\rm i}}\; \mathrm{d}\tau'.
\end{equation}
We now invoke the stationary phase approximation, so that $\Phi(\tau)\simeq
\Phi(\tau_0) + (1/2)\ddot\Phi (\tau_0)(\tau - \tau_0)^2$, where $\tau_0$ is
determined by $\dot\Phi = 0$, occurring when $\eta_0 = \cos \theta_i /\cos I$.
We then find, for $\tau \gg \tau_0$,
\begin{equation}
    S(\tau)e^{-i\Phi(\tau)} - S(\tau_{\rm i})
        \simeq -i \eta(\tau_0)\sin I \cos \theta_{\rm i}\,e^{-i\Phi(\tau_0)}
            \sqrt{\frac{2\pi}{i \ddot\Phi(\tau_0)}}.
\end{equation}
Using $\dot\eta = -\epsilon \eta$ [Eq.~\eqref{eq:deta_dt}] and
$\ddot\Phi(\tau_0) = \dot\eta(\tau_0)\cos I = -\epsilon \cos\theta_i$,
we have
\begin{equation}
  S(\tau)e^{-i\Phi(\tau)} - S(\tau_{\rm i})
        \simeq -i^{3/2}\tan I (\cos\theta_{\rm i})^{3/2} e^{-i\Phi(\tau_0)}
            \sqrt{\frac{2\pi}{\epsilon}}.
\end{equation}
The final obliquity $\theta_{\rm f}$ is then given by
\begin{equation}
    \sin\theta_{\rm f}\simeq \abs{
        \sin\theta_i+ e^{-i\varphi_0}\tan I (\cos\theta_{\rm i})^{3/2}
            \sqrt{\frac{2\pi}{\epsilon}}},
\label{eq:thetaf}
\end{equation}
where $\varphi_0=\Phi(\tau_0)+\pi/4$ is a constant phase. If the initial
obliquity is much smaller than the final obliquity ($\sin\theta_{\rm i} \ll
\sin\theta_{\rm f}$), we obtain
\begin{equation}
    \sin\theta_{\rm f}\simeq \sqrt{\frac{2\pi}{\epsilon}}
        \tan I (\cos\theta_{\rm i})^{3/2}.
            \label{eq:nonad_q_f}
\end{equation}
This expression is valid only if $\cos \theta \approx \cos \theta_{\rm i}$
throughout the evolution. This corresponds to the limit where $\theta_{\rm f}$
is not much larger than $\theta_{\rm i}$, which requires $\epsilon$ not to be
too small. Numerically, this is consistent with the system being in the
nonadiabatic regime $\epsilon \gtrsim \epsilon_{\rm c}$ (see
Fig.~\ref{fig:nonad_3_scan}).

The above calculation applies for a specific initial $\theta_{\rm i}$, but, as
discussed at the beginning of Section~\ref{ss:ad_ensemble}, the initial spin
orientation is more appropriately described by $\theta_{\rm sd, i}$ since
$\eta_i\gg 1$. The correct way to predict the final obliquity for a given
$\theta_{\rm sd, i}$ using Eq.~\eqref{eq:thetaf} is somewhat subtle but yields
good agreement with numerical results.

First consider the case with $\theta_{\rm sd, i} = 0$. This corresponds to a
well-defined initial obliquity $\theta_i = I$ (more precisely, the initial
condition is CS2). The final obliquity in this case, denoted
$\theta_{\rm 0f}$, is given by
\begin{align}
    \sin\theta_{\rm 0f} &\simeq \sin I \abs{1 + e^{-i\varphi_0}
            \sqrt{\frac{2\pi \cos I}{\epsilon}}},\nonumber\\
        &\approx \sin I \sqrt{\frac{2\pi \cos I}{\epsilon}},
            \label{eq:theta_2f}
\end{align}
where the second equality assumes $\sqrt{2\pi/\epsilon}\gg 1$.
Fig.~\ref{fig:nonad_3_scan} shows the final obliquity as function of $\epsilon$
for $\theta_{\rm sd,i} = 0$ and $I = 5^\circ$. We see that the agreement between
the numerical results and Eq.~\eqref{eq:theta_2f} is excellent. For $\epsilon
\ll \epsilon_{\rm c}$, we find $\theta_{\rm f} \simeq 90^\circ$, in agreement
with the result of adiabatic evolution (see Fig.~\ref{fig:ad_ensemble}).

\begin{figure}
    \centering
    \includegraphics[width=0.47\textwidth]{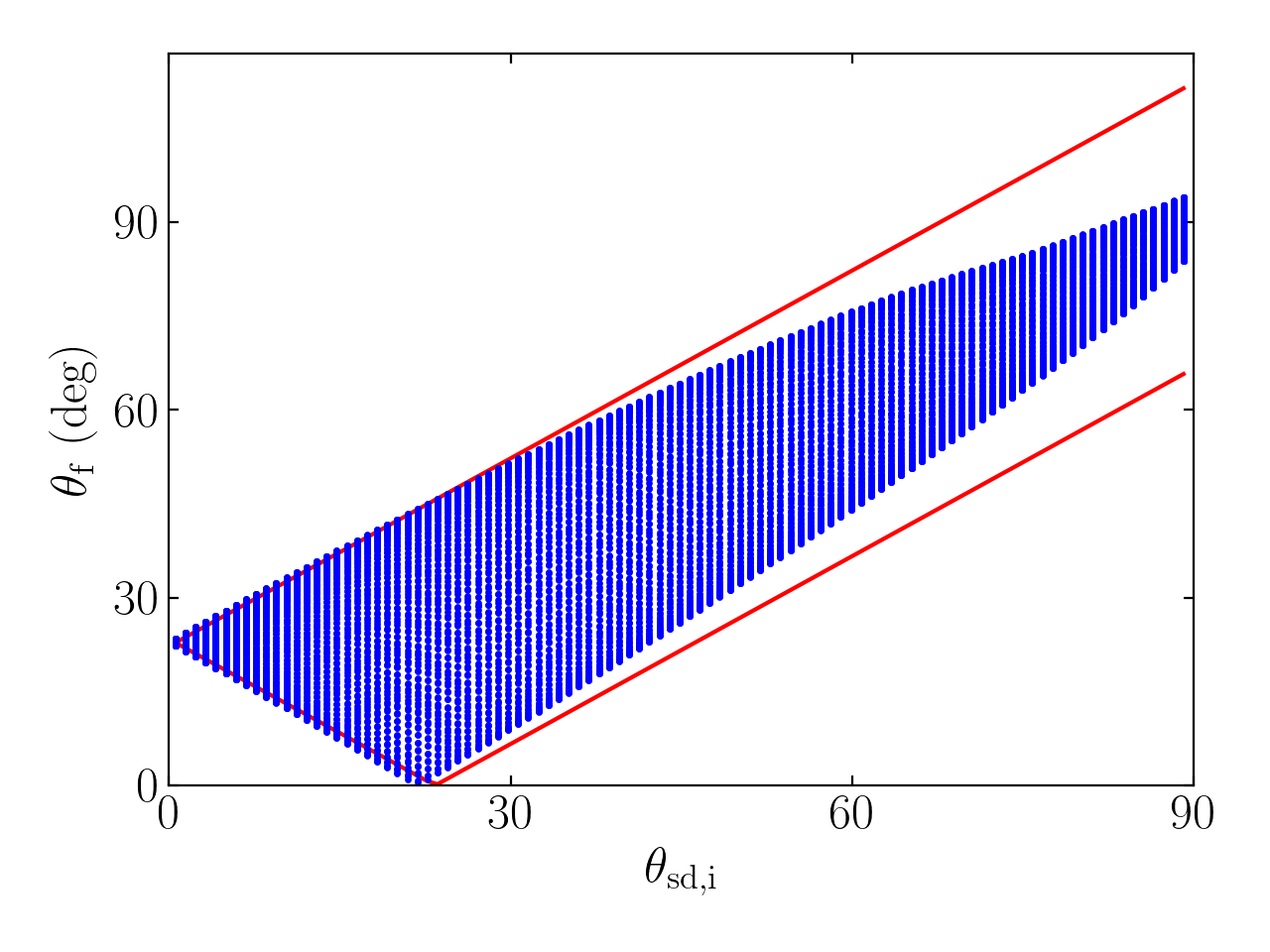}
    \caption{Final obliquity $\theta_{\rm  f}$ vs $\theta_{\rm sd, i}$ for
    $I=5^\circ$ and $\epsilon = 0.3$ (firmly in the nonadiabatic regime). The
    blue dots represent numerical results, and the two red lines show the
    analytical lower and upper bounds given by
    Eq.~\eqref{eq:nonad_q_f_dist}.}\label{fig:nonad_3_ensemble}
\end{figure}

\begin{figure}
    \centering
    \includegraphics[width=0.47\textwidth]{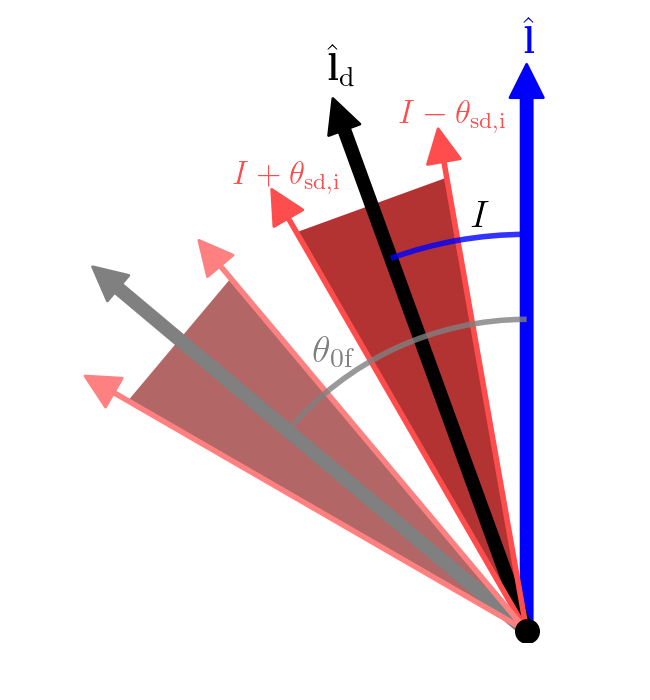}
    \caption{Schematic picture for understanding nonadiabatic obliquity
    evolution when $\theta_{\rm sd, i} > 0$ assuming $\eta_{\rm i} \gg 1$. The
    figure shows a projection onto the plane containing both $\uv{l}$ and
    $\uv{l}_{\rm d}$. When $\theta_{\rm sd, i} = 0$, the initial spin vector
    points along $\uv{l}_{\rm d}$ and evolves into the final spin vector (grey),
    which has obliquity $\theta_{\rm 0f}$ [Eq.~\eqref{eq:theta_2f}]. When
    $\theta_{\rm sd, i} \neq 0$, the set of initial conditions for the spin
    vector forms a cone (solid red area) centered on $\uv{l}_{\rm d}$ with
    opening half-angle $\theta_{\rm sd, i}$. Under nonadiabatic evolution, the
    set of final spin vectors forms a new cone, still with opening half-angle
    $\theta_{\rm sd, i}$, centered on $\theta_{\rm 0f}$ (light red
    area).}\label{fig:2nonadrot}
\end{figure}

When $\theta_{\rm sd, i}\neq 0$, we find that the final obliquity $\theta_{\rm
f}$ spans a range of values for a given $\theta_{\rm sd, i}$, as can be
seen in Fig.~\ref{fig:nonad_3_ensemble}. The range can be described by
\begin{equation}
    \abs{\theta_{\rm 0f} - \theta_{\rm sd, i}}
        \lesssim \theta_{\rm f}
        \lesssim \theta_{\rm 0f} + \theta_{\rm sd, i}
    .\label{eq:nonad_q_f_dist}
\end{equation}
Eq.~\eqref{eq:nonad_q_f_dist} can be understood as follows (see
Fig.~\ref{fig:2nonadrot}). In the beginning ($\eta = \eta_{\rm i} \gg 1$), the
initial spin vector precesses around $\uv{l}_{\rm d}$ on a cone with opening
half-angle $\theta_{\rm sd, i}$ (more precisely, the cone is centered on CS2,
which coincides with $\uv{l}_{\rm d}$ as $\eta_{\rm i} \to \infty$). Note that
for $\eta \gg 1$, the adiabaticity condition is easily satisfied: using
$\theta_2 \simeq I + \eta^{-1}\sin I \cos I$ (see
Section~\ref{s:local_dynamics}), Eq.~\eqref{eq:w_lib} gives $\omega_{\rm lib}
\simeq \eta$ while Eq.~\eqref{eq:dq2dt} gives $\abs{\rdil{\theta_2}{\tau}}
\simeq \p{\epsilon / \eta} \sin I \cos I \ll \omega_{\rm lib}$. As $\eta$
decreases, the system will transition from being adiabatic to being
nonadiabatic, since $\epsilon \gtrsim \epsilon_{\rm c}$. The evolution of the
system can thus be decomposed into two phases: (i) when the evolution is
adiabatic, the spin vector will precess around the slowly-moving CS2; (ii) when
the evolution becomes nonadiabatic, the spin vector stops tracking the
quickly-evolving CS2. During the adiabatic evolution of phase (i), the angle
between CS2 and the spin vector is approximately unchanged due to conservation
of phase space area\footnote{ This approximation assumes sufficiently small
$\theta_{\rm sd, i}$ such that libration about CS2 remains approximately
circular throughout phase (i) (initially, when $\eta \to \infty$, all librations
are circular about $\uv{l}_{\rm d}$). This assumption breaks down when
$\theta_{\rm sd, i}$ is sufficiently large that librating orbits become
non-circular as $\eta$ decreases before the end of phase (i)
(Fig.~\ref{fig:eq_1contours} illustrates that the libration cycles farther from
CS2, corresponding to a larger $\theta_{\rm sd, i}$, are less circular for a
given $\eta$).
This causes the deviation of the numerical results in
Fig.~\ref{fig:nonad_3_ensemble} from Eq.~\eqref{eq:nonad_q_f_dist} for
$\theta_{\rm sd, i} \gtrsim 45^\circ$. }.
Once the evolution enters phase (ii), the precession axis
quickly (on timescale $\ll 1 / \omega_{\rm lib}$) changes to $\uv{l}$ (as $\eta$
decreases to $\eta_{\rm f} \ll 1$). Precession about $\uv{l}$ does not change the
obliquity, so the range of obliquities at the end of phase (i) is frozen in as
the range of final obliquities. We refer to this two-phase
evolution as \emph{partial adiabatic resonance advection}.

Fig.~\ref{fig:nonad_3_ensemble} shows the numerical result of $\theta_{\rm f}$
vs $\theta_{\rm sd,i}$ for $I =5^\circ$ and $\epsilon = 0.3$. We see that
Eq.~\eqref{eq:nonad_q_f_dist} provides good lower and upper bounds of the final
obliquity for $\theta_{\rm sd,i} \lesssim 45^\circ$ (see footnote 3).


\section{Summary}\label{s:disk}

In this paper, we have studied the excitation of planetary obliquities due to
gravitational interaction with an exterior, dissipating (mass-losing)
protoplanetary disk. Obliquity excitation occurs as the system passes through a
secular resonance between spin precession and orbital (nodal) precession. This
scenario was recently studied by \citet{millholland_disk}, who focused on the
special case of small initial obliquities.  In contrast, we consider arbitrary
initial misalignment angles in this paper, motivated by the fact that planet
formation through core accretion can lead to a wide range of initial spin
orientations.  We present our result as a mapping from $\theta_{\rm sd, i}$ to
$\theta_{\rm f}$, where $\theta_{\rm sd, i}$ is the initial misalignment angle
between the planet's spin axis and the disk's orbital angular momentum axis, and
$\theta_{\rm f}$ is the final planetary obliquity. We have derived analytical
results that capture the behavior of this mapping in both the adiabatic and
nonadiabatic limits:
\begin{enumerate}
    \item In the adiabatic limit (i.e.\ the disk dissipates at a
        sufficiently slow rate), we reproduce the known result $\theta_{\rm f}
        \simeq 90^\circ$ for $\theta_{\rm sd, i} \simeq 0$. We demonstrate (via
        numerical calculation and analytical argument) the dual-valued behavior
        of $\theta_{\rm f}$ for nonzero $\theta_{\rm sd, i}$ (see
        Fig.~\ref{fig:ad_ensemble}). We show for the first time
        that both the final $\theta_{\rm f}$ values and the probabilities of
        achieving each value can be understood analytically
        via careful accounting of adiabatic invariance and separatrix crossing
        dynamics.

    \item As the disk dissipates more rapidly, the adiabatic condition
        [Eq.~\eqref{eq:ad_constr}] breaks down, we find that a broad range of
        final obliquities can be reached for a given $\theta_{\rm sd, i}$ (see
        Fig.~\ref{fig:nonad_3_ensemble}). We understand this
        result via the novel concept of partial adiabatic resonance advection
        and provide an analytical expression of the bounds on $\theta_{\rm f}$
        in Eq.~\eqref{eq:nonad_q_f_dist}.
\end{enumerate}

As noted in Section~\ref{s:intro}, while in this paper we have examined a
specific scenario of generating/modifying planetary obliquities from planet-disk
interactions, the dynamical problem have studied is more general
\citep{colombo1966,peale1969,peale1974possible,ward1975tidal,henrard1987}. Our
work goes beyond these previous works and provides the most general solution to
the evolution of ``Colombo's top'' as the system evolves from the ``weak
spin-orbit coupling'' regime ($\eta\gg 1$) to the ``strong spin-orbit coupling''
regime ($\eta\ll 1$).  The new analytical results presented in this paper can be
adapted to other applications.

Concerning the production of planetary obliquity with a
dissipating disk, when there are multiple planets in a system, the nodal
precession rate $g$ for the planet of interest never decays below the
Laplace-Lagrange rate driven by planet-planet secular interactions
\citep{millholland_disk}. Therefore, $\eta$ has a minimum value at late times.
This does not affect the methodology of our analysis, but can affect the
detailed results. For example, the adiabaticity criterion must be modified
slightly as $\rdil{\ln \eta}{t}$ is no longer constant but asymptotes to zero as
$\eta$ decreases; the planet may never undergo separatrix crossing if their
$\eta_{\star}$ (which depends on $\theta_{\rm sd, i}$) in the absence of the
companions is too small; the planetary obliquity will oscillate even when the
disk has fully evaporated (as $\uv{l}$ is no longer constant). The spin dynamics
can be even more complex if the two planets are in mean motion resonance
\citep[e.g.][]{millholland2019obliquity}.

\section*{Acknowledgements}

We thank the anonymous reviewer for detailed comments that
improved this work and Kassandra Anderson for discussion and assistance in the
early phase of this work. DL thanks the Dept.\ of Astronomy and the Miller
Institute for Basic Science at UC Berkeley for hospitality while part of this
work was carried out. This work has been supported in part by the NSF grant
AST-17152 and NASA grant 80NSSC19K0444. YS is supported by the NASA FINESST
grant 19-ASTRO19-0041.

\bibliography{Su_disk_disp}

\begin{thebibliography}{}
\expandafter\ifx\csname natexlab\endcsname\relax\def\natexlab#1{#1}\fi
\providecommand{\url}[1]{\href{#1}{#1}}
\providecommand{\dodoi}[1]{doi:~\href{http://doi.org/#1}{\nolinkurl{#1}}}
\providecommand{\doeprint}[1]{\href{http://ascl.net/#1}{\nolinkurl{http://ascl.net/#1}}}
\providecommand{\doarXiv}[1]{\href{https://arxiv.org/abs/#1}{\nolinkurl{https://arxiv.org/abs/#1}}}

\bibitem[{Adams {et~al.}(2019)Adams, Millholland, \&
  Laughlin}]{millholland_signatures}
Adams, A.~D., Millholland, S., \& Laughlin, G.~P. 2019, arXiv preprint
  arXiv:1906.07615

\bibitem[{Anderson \& Lai(2018)}]{anderson2018teeter}
Anderson, K.~R., \& Lai, D. 2018, Monthly Notices of the Royal Astronomical
  Society, 480, 1402

\bibitem[{Batygin \& Adams(2013)}]{batygin2013magnetic}
Batygin, K., \& Adams, F.~C. 2013, The Astrophysical Journal, 778, 169

\bibitem[{Benz {et~al.}(1989)Benz, Slattery, \& Cameron}]{benz1989tilting}
Benz, W., Slattery, W., \& Cameron, A. 1989, Meteoritics, 24, 251

\bibitem[{Bryan {et~al.}(2018)Bryan, Benneke, Knutson, Batygin, \&
  Bowler}]{bryan2018constraints}
Bryan, M.~L., Benneke, B., Knutson, H.~A., Batygin, K., \& Bowler, B.~P. 2018,
  Nature Astronomy, 2, 138

\bibitem[{Bryan {et~al.}(2020)Bryan, Chiang, Bowler, Morley, Millholland,
  Blunt, Ashok, Nielsen, Ngo, Mawet, {et~al.}}]{bryan2020obliquity}
Bryan, M.~L., Chiang, E., Bowler, B.~P., {et~al.} 2020, The Astronomical
  Journal, 159, 181

\bibitem[{{Colombo}(1966)}]{colombo1966}
{Colombo}, G. 1966, SAO Special Report, 203

\bibitem[{Correia {et~al.}(2003)Correia, Laskar, \& de~Surgy}]{correia2003long}
Correia, A.~C., Laskar, J., \& de~Surgy, O.~N. 2003, Icarus, 163, 1

\bibitem[{Dones \& Tremaine(1993)}]{dones1993does}
Dones, L., \& Tremaine, S. 1993, Science, 259, 350

\bibitem[{Fabrycky {et~al.}(2007)Fabrycky, Johnson, \&
  Goodman}]{fabrycky_otides}
Fabrycky, D.~C., Johnson, E.~T., \& Goodman, J. 2007, The Astrophysical
  Journal, 665, 754

\bibitem[{Hamilton \& Ward(2004)}]{ward2004II}
Hamilton, D.~P., \& Ward, W.~R. 2004, The Astronomical Journal, 128, 2510

\bibitem[{Henrard(1982)}]{henrard1982}
Henrard, J. 1982, Celestial Mechanics and Dynamical Astronomy, 27, 3

\bibitem[{Henrard \& Murigande(1987)}]{henrard1987}
Henrard, J., \& Murigande, C. 1987, Celestial Mechanics, 40, 345

\bibitem[{Inamdar \& Schlichting(2015)}]{inamdar2015formation}
Inamdar, N.~K., \& Schlichting, H.~E. 2015, Monthly Notices of the Royal
  Astronomical Society, 448, 1751

\bibitem[{Izidoro {et~al.}(2017)Izidoro, Ogihara, Raymond, Morbidelli, Pierens,
  Bitsch, Cossou, \& Hersant}]{izidoro2017breaking}
Izidoro, A., Ogihara, M., Raymond, S.~N., {et~al.} 2017, Monthly Notices of the
  Royal Astronomical Society, 470, 1750

\bibitem[{Korycansky {et~al.}(1990)Korycansky, Bodenheimer, Cassen, \&
  Pollack}]{korycansky1990one}
Korycansky, D., Bodenheimer, P., Cassen, P., \& Pollack, J. 1990, Icarus, 84,
  528

\bibitem[{Lai(2014)}]{lai2014star}
Lai, D. 2014, Monthly Notices of the Royal Astronomical Society, 440, 3532

\bibitem[{Lainey(2016)}]{lainey2016quantification}
Lainey, V. 2016, Celestial Mechanics and Dynamical Astronomy, 126, 145

\bibitem[{Laskar \& Robutel(1993)}]{laskar1993chaotic}
Laskar, J., \& Robutel, P. 1993, Nature, 361, 608

\bibitem[{Lissauer {et~al.}(1997)Lissauer, Berman, Greenzweig, \&
  Kary}]{lissauer1997accretion}
Lissauer, J.~J., Berman, A.~F., Greenzweig, Y., \& Kary, D.~M. 1997, Icarus,
  127, 65

\bibitem[{Miguel \& Brunini(2010)}]{miguel2010planet}
Miguel, Y., \& Brunini, A. 2010, Monthly Notices of the Royal Astronomical
  Society, 406, 1935

\bibitem[{Millholland \& Batygin(2019)}]{millholland_disk}
Millholland, S., \& Batygin, K. 2019, The Astrophysical Journal, 876, 119

\bibitem[{Millholland \& Laughlin(2018)}]{millholland_wasp12b}
Millholland, S., \& Laughlin, G. 2018, The Astrophysical Journal Letters, 869,
  L15

\bibitem[{Millholland \& Laughlin(2019)}]{millholland2019obliquity}
---. 2019, Nature Astronomy, 3, 424

\bibitem[{Morbidelli {et~al.}(2012)Morbidelli, Tsiganis, Batygin, Crida, \&
  Gomes}]{morbidelli_gi}
Morbidelli, A., Tsiganis, K., Batygin, K., Crida, A., \& Gomes, R. 2012,
  Icarus, 219, 737

\bibitem[{Ohno \& Zhang(2019)}]{ohno_infer_obl}
Ohno, K., \& Zhang, X. 2019, The Astrophysical Journal, 874, 2

\bibitem[{Peale(1969)}]{peale1969}
Peale, S.~J. 1969, The Astronomical Journal, 74, 483

\bibitem[{Peale(1974)}]{peale1974possible}
---. 1974, The Astronomical Journal, 79, 722

\bibitem[{Rogoszinski \& Hamilton(2019)}]{hamilton_tilting_ice}
Rogoszinski, Z., \& Hamilton, D.~P. 2019, arXiv preprint arXiv:1908.10969

\bibitem[{Safronov \& Zvjagina(1969)}]{original_gi}
Safronov, V., \& Zvjagina, E. 1969, Icarus, 10, 109

\bibitem[{Seager \& Hui(2002)}]{seager2002constraining}
Seager, S., \& Hui, L. 2002, The Astrophysical Journal, 574, 1004

\bibitem[{Snellen {et~al.}(2014)Snellen, Brandl, de~Kok, Brogi, Birkby, \&
  Schwarz}]{snellen2014fast}
Snellen, I.~A., Brandl, B.~R., de~Kok, R.~J., {et~al.} 2014, Nature, 509, 63

\bibitem[{Touma \& Wisdom(1993)}]{touma1993chaotic}
Touma, J., \& Wisdom, J. 1993, Science, 259, 1294

\bibitem[{Vokrouhlick{\`y} \& Nesvorn{\`y}(2015)}]{vokrouhlicky2015tilting}
Vokrouhlick{\`y}, D., \& Nesvorn{\`y}, D. 2015, The Astrophysical Journal, 806,
  143

\bibitem[{Ward(1975)}]{ward1975tidal}
Ward, W.~R. 1975, The Astronomical Journal, 80, 64

\bibitem[{Ward \& Hamilton(2004)}]{ward2004I}
Ward, W.~R., \& Hamilton, D.~P. 2004, The Astronomical Journal, 128, 2501

\bibitem[{Zanazzi \& Lai(2018)}]{zanazzi2018planet}
Zanazzi, J., \& Lai, D. 2018, Monthly Notices of the Royal Astronomical
  Society, 478, 835

\end{thebibliography}
\bibliographystyle{aasjournal}

\appendix

\section{Cassini State Local Dynamics}\label{s:local_dynamics}

In this appendix, we linearize the equations of motion near each CS and
determine its stability. We derive the local libration frequency or growth
rate for perturbations around each CS\@.

\subsection{Canonical Equations of Motion and Solutions}\label{ss:canonical}

We adopt spherical coordinate system where $\uv{l} =
\uv{z}$ and $\theta, \phi$ are the polar and azimuthal angle of $\uv{s}$. We
choose $\uv{l}_{\rm d}$ at coordinates $\theta = I, \phi = \pi$ (see
Figs.~\ref{fig:cs_vecs} and~\ref{fig:cs_locs}). We use the convention
$\theta \in [0, \pi)$ 
and $\phi \in [0, 2\pi)$.

The equations of motion in $\p{\phi, \cos \theta}$ follow by
applying Hamilton's equations to the Hamiltonian [Eq.~\eqref{eq:H}]:
\begin{subequations}\label{se:H_eom}
    \begin{align}
        \rd{\phi}{t} = \pd{\mathcal{H}}{(\cos\theta)}
            &= -\cos\theta + \eta\p{\cos I + \sin I \cot \theta \cos \phi},
                \label{seq:H_eom_phi_t}\\
        \rd{(\cos \theta)}{t} = -\pd{\mathcal{H}}{\phi}
            &= -\eta \sin I \sin \theta \sin \phi.
                \label{seq:H_eom_mu_t}
    \end{align}
\end{subequations}
These agree with Eq.~\eqref{eq:dsdt_base}.

The CSs satisfy $\dot{\phi} = \dot{\theta} = 0$. For convenience, we give
approximate solutions for the CSs in the limits $\eta \ll 1$ and $\eta \gg 1$.
For $\eta \ll 1$:
\begin{itemize}
    \item CS1: $\phi_1 = 0$, $\theta_1 \simeq \eta \sin I$.
    \item CS2: $\phi_2 = \pi$, $\theta_2 \simeq \pi / 2 - \eta \cos I$.
    \item CS3: $\phi_3 = 0$, $\theta_3 \simeq \pi - \eta \sin I$.
    \item CS4: $\phi_4 = 0$, $\theta_4 \simeq \pi / 2 - \eta \cos I$.
\end{itemize}
For $\eta \gg 1$, only CS2 and CS3 exist and are given by:
\begin{itemize}
    \item CS2: $\phi_2 = \pi$, $\theta_2 \simeq I + \eta^{-1}\sin I \cos I$.
    \item CS3: $\phi_3 = 0$, $\theta_3 \simeq \pi - I + \eta^{-1}\sin I \cos I$.
\end{itemize}
Note that in the convention of Fig.~\ref{fig:cs_locs}, CS1, CS3 and CS4 have
negative $\theta$ values since $\phi=0$.

\subsection{Stability and Frequency of Local Oscillations}\label{ss:eigens}

\begin{figure}
    \centering
    \includegraphics[width=0.47\textwidth]{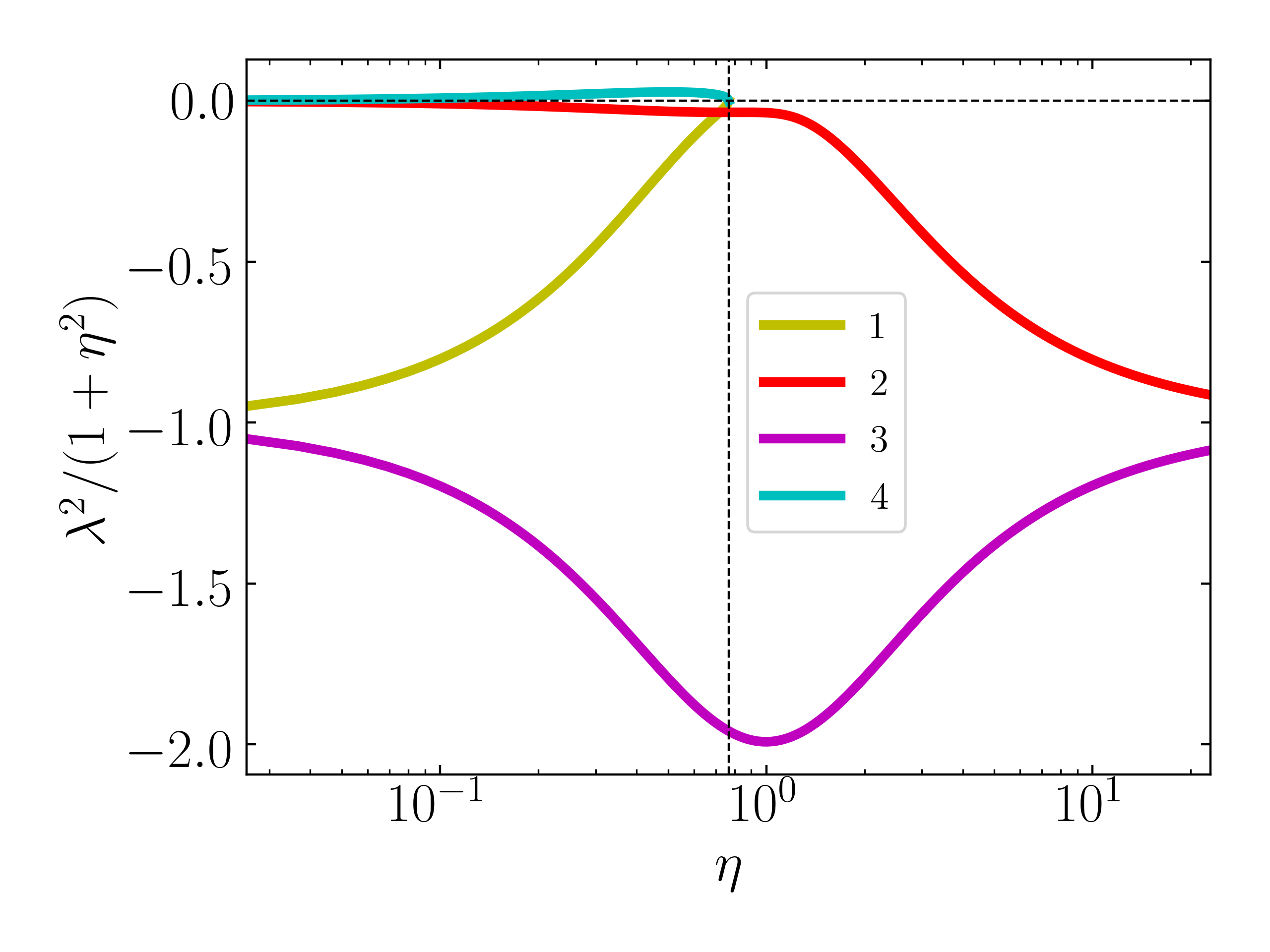}
    \caption{$\lambda^2$, given by Eq.~\eqref{eq:lambda2}, evaluated at each of
    the Cassini States. The vertical axis is rescaled for clarity. Note that CS4
    is unstable ($\lambda^2 > 0$) when it exists while all others are
    stable ($\lambda^2 < 0$). The thin horizontal dashed line is the instability
    boundary $\lambda^2 = 0$ while the thin vertical dashed line labels $\eta =
    \eta_{\rm c}$ [Eq.~\eqref{eq:etac}].}\label{fig:lambda2}
\end{figure}

To examine stability of each CS, we linearize Eqs.~\eqref{se:H_eom} about an
equilibrium located at $\phi_{\rm cs} = 0$ (CS 1, 3, 4) or $\pi$ (CS2) but
arbitrary $\theta_{\rm cs}$. Setting $\phi = \phi_{\rm cs} + \delta \phi, \theta
= \theta_{\rm cs} + \delta \theta$ yields
\begin{subequations}\label{se:H_eom_lin}
    \begin{align}
        \rd{\delta \phi}{t} &= \sin \theta_{\rm cs} \delta \theta
            \mp \eta \frac{\sin I}{\sin^2\theta_{\rm cs}} \delta \theta,\\
        \rd{\delta \theta}{t} &= \pm \eta \sin I \delta \phi,
    \end{align}
\end{subequations}
where the upper sign corresponds to $\phi_{\rm cs} = 0$. Eliminating $\delta
\theta$ gives
\begin{equation}
    \rtd{\delta \phi}{t} \equiv \lambda^2\delta \phi,\label{eq:lambda2}
\end{equation}
where
\begin{equation}
    \lambda^2 \equiv \p{\sin \theta_{\rm cs}
        \mp \eta \sin I\csc^2\theta}\p{\pm \eta \sin I}.
\end{equation}
A plot of $\lambda^2$ for each of the CSs is given in Fig.~\ref{fig:lambda2}.
It is clear that CS4 is unstable while the other three are stable. The local
libration frequency for these stable CSs is simply $\omega_{\rm lib} =
\sqrt{-\lambda^2}$.

\section{Approximate Adiabatic Evolution}\label{s:ad_approx}

In this appendix, we will use approximations valid for small $\eta$ to derive
the explicit analytic expressions for the final obliquities at small
$\theta_{\rm sd, i}$ and the associated probabilities for the II $\to$ I and II
$\to$ III tracks. These are the only possible tracks for small $\eta$.

We first seek a simple parameterization for the separatrix, the level curve of
the Hamiltonian intersecting the unstable equilibrium CS4. Points along the
separatrix, parameterized by $\p{\phi, \theta_{\rm sep}(\phi)}$, satisfy
$\mathcal{H}\p{\phi, \theta_{\rm sep}(\phi)} = \mathcal{H}\p{\phi_4, \theta_4}$
where $\phi_4$ and $\theta_4$ are given in Appendix~\ref{ss:canonical}. We
obtain two solutions for $\theta_{\rm sep}$, given to leading order in $\eta$
by:
\begin{equation}
    \cos \theta_{\rm sep}(\phi) \approx \cos \theta_4 \pm
        \sqrt{2\eta \sin I\p{1 - \cos \phi}}.
\end{equation}
These two solutions parameterize the two legs of the separatrix. Integration of
the phase area enclosed by the separatrix yields then
\begin{equation}
    \mathcal{A}_{\rm II}(\eta) \approx 16\sqrt{\eta \sin I}.\label{eq:a_approx}
\end{equation}
We can now compute the final obliquities and their associated probabilities for
each track as follows:
\begin{enumerate}
    \item For a given $\theta_{\rm sd, i}$, we know that if $\eta \to \infty$
        then the trajectory executes simple libration about $\uv{l}_{\rm d}$,
        and so $A = 2\pi\p{1 - \cos \theta_{\rm sd, i}} \approx \pi \theta_{\rm
        sd, i}^2$. This then implies $\eta_\star$ must be the solution to
        $\mathcal{A}_{\rm II}(\eta_\star) = A$, or
        \begin{equation}
            \eta_\star \approx \p{\frac{2\pi\p{1 - \cos \theta_{\rm sd,i}}}{
                        16}}^2 \frac{1}{\sin I}
                    \approx \p{\frac{\pi \theta_{\rm sd,
                    i}^2}{16}}^2\frac{1}{\sin I}.
        \end{equation}

    \item Upon separatrix encounter, a transition to either zone I or zone
        III occurs. These can be calculated to have the associated probabilities
        [using the approximate area Eq.~\eqref{eq:a_approx} and
        Eqs.~\eqref{eq:henrard_hop}]
        \begin{subequations}
            \begin{align}
                \Pr\p{\rm II \to I} &\approx \frac{2\pi
                    \eta_{\star} \cos I + 4\sqrt{\eta_{\star}\sin
                    I}}{8\sqrt{\eta_{\star}\sin I}},\\
                \Pr\p{\rm II \to III} &\approx \frac{-2\pi
                    \eta_{\star} \cos I + 4\sqrt{\eta_{\star}\sin
                    I}}{8\sqrt{\eta_{\star}\sin I}}.
            \end{align}
        \end{subequations}

    \item Upon a transition to zone I or zone III, the final obliquity can be
        predicted by observing the final adiabatic invariant $A_{\rm f} =
        -\mathcal{A}_{\rm I}(\eta_\star)$ in the zone I case and $A_{\rm f} =
        \mathcal{A}_{\rm I}(\eta_\star) + \mathcal{A}_{\rm I}I(\eta_\star)$ in
        the zone III case. As $\eta
        \to 0$, these correspond to obliquities
        \begin{subequations}\label{se:q_f_approx}
            \begin{align}
                \p{\cos \theta_{\rm f}}_{\rm II \to I} &\approx
                    \p{\frac{\pi \theta_{\rm sd, i}^2}{16}}^2 \cot I
                        + \frac{\theta_{\rm sd, i}^2}{4},\\
                \p{\cos \theta_{\rm f}}_{\rm II \to III} &\approx
                    \p{\frac{\pi \theta_{\rm sd, i}^2}{16}}^2 \cot I
                        - \frac{\theta_{\rm sd, i}^2}{4}.
            \end{align}
        \end{subequations}
        These are the black dotted lines overplotted in
        Fig.~\ref{fig:ad_ensemble}.
\end{enumerate}

\end{document}